\documentclass[preprint2]{emulateapj}

\shorttitle{Coronagraphy for Arbitrarily shaped Apertures}
\shortauthors{Guyon}
\usepackage{graphicx}
\usepackage[mathscr]{eucal}
\begin{document}

\title{High Performance Lyot and PIAA Coronagraphy for Arbitrarily shaped Telescope Apertures}

\author{Olivier Guyon}
\affil{Steward Observatory, University of Arizona, Tucson, AZ 85721, USA}
\affil{National Astronomical Observatory of Japan, Subaru Telescope, Hilo, HI 96720}
\email{guyon@naoj.org}

\author{Philip M. Hinz}
\affil{Steward Observatory, University of Arizona, Tucson, AZ 85721, USA}

\author{Eric Cady}
\affil{Jet Propulsion Laboratory, 4800 Oak Grove Drive, Pasadena, CA 91109, USA}

\author{Ruslan Belikov}
\affil{NASA Ames Research Center, Moffett Field, CA 94035, USA}

\author{Frantz Martinache}
\affil{National Astronomical Observatory of Japan, Subaru Telescope, Hilo, HI 96720}

\begin{abstract}
Two high performance coronagraphic approaches compatible with segmented and obstructed telescope pupils are described. Both concepts use entrance pupil amplitude apodization and a combined phase and amplitude focal plane mask to achieve full coronagraphic extinction of an on-axis point source. While the first concept, named Apodized Pupil Complex Mask Lyot Coronagraph (APCMLC), relies on a transmission mask to perform the pupil apodization, the second concept, named Phase-Induced Amplitude Apodization complex mask coronagraph (PIAACMC), uses beam remapping for lossless apodization. Both concepts theoretically offer complete coronagraphic extinction (infinite contrast) of a point source in monochromatic light, with high throughput and sub-$\lambda$/D inner working angle, regardless of aperture shape. The PIAACMC offers nearly 100\% throughput and approaches the fundamental coronagraph performance limit imposed by first principles. The steps toward designing the coronagraphs for arbitrary apertures are described for monochromatic light. Designs for the APCMLC and the higher performance PIAACMC are shown for several monolith and segmented apertures, such as the apertures of the Subaru Telescope, Giant Magellan Telescope (GMT), Thirty Meter Telescope (TMT), the European Extremely Large Telescope (E-ELT) and the Large Binocular Telescope (LBT). Performance in broadband light is also quantified, suggesting that the monochromatic designs are suitable for use in up to 20\% wide spectral bands for ground-based telescopes.
\end{abstract}

\section{Introduction}
\label{sec:intro}

Direct imaging of exoplanets with ground-based telescopes is becoming possible thanks to advances in adaptive optics, as demonstrated by several recent direct imaging exoplanet discoveries \citep{2010Sci...329...57L,2008Sci...322.1348M,2013ApJ...763L..32C}. While current ground-based instruments are most sensitive to relatively massive and young planets at large angular separation (typically beyond a few tenths of an arcsecond), recent developments in coronagraphic techniques, ``extreme'' Adaptive Optics and calibration techniques are pushing detection limits deeper in contrast and closer in angular separation, soon providing access to the planet-rich inner parts of planetary systems \citep{2008SPIE.7015E..31M,2008SPIE.7014E..41B,2009SPIE.7440E..20M,2011ApJ...729..132C}. Direct imaging of the inner part (1 to 5 AU) of young planetary systems is of especially high scientific value to constrain and understand planetary systems formation and evolution near the habitable zone, and requires the combination of an efficient coronagraph offering small inner working angle and a high level of wavefront correction and calibration.

High contrast imaging from space allows access to considerably better contrast than possible with ground-based telescopes, thanks to the absence of atmospheric turbulence. Laboratory coronagraphy systems have demonstrated that raw contrasts of about 1e-9 can be achieved in a stable environment with a deformable mirror and a coronagraph (see for example \cite{2007Natur.446..771T}). At such high contrast, coronagraphic imaging can allow characterization of potentially habitable planets through spectroscopy from space \citep{2009arXiv0911.3200L}. 

While most high performance coronagraphs are designed for unobstructed circular pupils, current and future large ground-based telescopes are centrally obscured, and also segmented above 8.4-m diameter. Future large space-based telescopes will also likely be centrally obscured and/or segmented, although a telescope dedicated to high contrast imaging could be built off-axis if required for coronagraphy \citep{2009arXiv0911.3200L}. The scientific return of an exoplanet direct imaging mission or instrument is a steep function of telescope diameter: larger telescopes allow access to exoplanets at smaller angular separations, which are brighter in reflected light (apparent luminosity scales as inverse square of angular separation in reflected light), more numerous (the number of planets of a given type accessible with a telescope scales as the third power of telescope diameter), and more relevant to exoplanet systems habitability than widely separated planets. Larger ground-based telescope size also allows higher contrast observation by better concentrating planet light over the speckle halo background, and the gain in collecting area enables spectroscopic characterization. It is therefore essential to identify and develop coronagraph concepts which can deliver high performance on centrally obscured and/or segmented apertures.

Coronagraph designs for centrally obscured and/or segmented apertures have been proposed by several authors, offering a wide range of solutions and approaches: 
\begin{itemize}
\item{{\bf Lyot-type coronagraphs with amplitude masks.} Most studies of coronagraph designs for obscured and/or segmented apertures considered Lyot-type coronagraph optimized for high contrast by either apodization of the entrance pupil (APLC concept introduced by \cite{2003AA...397.1161S}) or apodization of the focal plane mask (Band-limited coronagraph concept introduced by \cite{2002ApJ...570..900K}). For the apodized pupil Lyot coronagraph (APLC), \cite{2005ApJ...618L.161S} and \cite{2009ApJ...695..695S} showed that the entrance pupil apodizer can be optimized for centrally obscured pupils. Using this technique, \cite{2007AA...474..671M} studied the APLC for ELTs, finding high throughput solutions offering better than 1e-5 contrast at and beyond 3 $\lambda$/D separation. \cite{2010AA...519A..61M} proposed using a multistage apodized pupil Lyot coronagraph (APLC) to mitigate central obstruction limitations. While central obstruction can be mitigated in the Lyot-type coronagraph design, \cite{2005ApJ...633..528S,2005ApJ...626L..65S} showed that spiders and gaps in APLC and band-limited Lyot coronagraphs diffract light within the geometrical aperture, making it impossible to achieve very high contrast on segmented apertures. Moderate-contrast band-limited Lyot coronagraphs have been designed for the NIRCAM instrument \citep{2009SPIE.7440E..28K} on the James Webb Space Telescope, but the aggressive Lyot stop designs, which remove much of the residual diffraction from the secondary structure and the segmented primary, come at the cost of significant companion throughput.}
\item{{\bf Phase mask coronagraphs.} Coronagraphs using phase focal plane masks are also affected by central obscuration and spiders/gaps. \cite{2003SPIE.4860..171L} showed that the 4-quadrant phase mask can only achieve full coronagraphic suppression on unobstructed pupils free of gaps or spiders, as any obscuration diffracts light outward in the Lyot plane. The optical vortex coronagraph is similarly affected by obscurations, although \cite{2011OptL...36.1506M} showed that central obstruction can be mitigated by a dual-stage approach, where the second stage rejects most of the light diffracted by the central obstruction. For both the vortex and the 4 quadrant phase mask coronagraphs, no solution has been found to eliminate the light diffracted by spiders and gaps. }
\item{{\bf Shaped Apertures.} For moderate contrast level and relatively large IWA, shaped apertures can be designed for centrally obscured and segmented pupils. \cite{2006PASJ...58..627T} designed shaped apertures delivering 1e-7 contrast at 4 $\lambda$/D. Similarly, \cite{2011arXiv1108.4050C} showed that shaped pupil can be designed for ~1e-6 contrast and $\approx 4 \lambda$/D inner working angle for a variety of centrally obscured and segmented apertures. }
\end{itemize}

A different approach to this problem is to remap the entrance aperture to remove central obstruction and/or spiders. \cite{2005PASP..117..295M} propose a 2-mirror system to remove central obstruction and spiders for a four-quadrant coronagraph. \cite{2006PASP..118..860G} propose a high efficiency nulling coronagraph concept adapted to central obstruction and spiders by performing destructive interferences between pairs of unobstructed off-axis subapertures. \cite{2009PASP..121.1232L} demonstrate that a prism-like transmissive device and aspheric optics can be used to remove both central obstruction and spiders from the Subaru Telescope pupil, theoretically allowing high performance coronagraphy with the full telescope aperture. These remapping solutions are complex, challenging to implement and align, and extremely sensitive to tip-tilt and stellar angular size \citep{2009ApJ...702..672C} at high contrast: when points on either size of an obstruction are brought next to each other in the remapped pupil, a small tip-tilt in the entrance beam leads to a phase discontinuity in the remapped beam. When due to finite stellar angular diameters, diffraction due to this discontinuity cannot be mitigated or controlled by wavefront control, as it is incoherent (opposite sides of the stellar disk produce diffracted light components of opposite signs). \cite{2007ApJ...658.1386S} chose to avoid entirely the problem by using an unobstructed 1.5-m diameter off-axis part of the 5-m Palomar telescope to perform high contrast imaging with the optical vortex coronagraph. While this allows the use of high performance coronagraphs designed for unobstructed apertures, the performance loss due to the use of an aperture considerably smaller than the full telescope is significant.

The solutions previously proposed to mitigate the effects of central obstruction, spiders and gaps are generally suitable for ground-based coronagraphy at a few $\lambda$/D IWA, with a raw contrast around $10^{-5}$, as reported by \cite{2008AA...492..289M} who performed a study of coronagraphic performance on ELTs including realistic assumptions on the level of residual wavefront error after an extreme-AO system. For most of the coronagraphs, central obstruction and spiders were found to have a major impact on performance, limiting the achievable contrast to $10^{-4}$ in the 1 to 4 $\lambda$/D separation range. The notable exceptions to this rule were the Achromatic Interfero Coronagraph \citep{2000A&AS..141..319B}, which is insensitive to centro-symmetric pupil features (such as a central obstruction or a set of four radial spiders at 90 deg), and the APLC, which could be designed to take into account central obstruction and was found to be quite robust to spiders at the $10^{-5}$ contrast level. The coronagraphs concepts for which ground-based designs compatible with central obscuration have been proposed (shaped aperture, APLC, band-limited Lyot coronagraph) are unfortunately not able to offer IWA less than $\approx 2 \lambda/D$, and also do not enable high contrast (approximately $10^{9}$) coronagraphy on centrally obscured or segmented apertures. 

The work presented in this paper is aimed at demonstrating that high performance coronagraphy is possible in monochromatic light on centrally obscured and/or segmented pupils for both ground-based and space-based telescopes. The Apodized Pupil Complex Mask Lyot Coronagraph (APCMLC) and Phase-Induced Amplitude Apodization complex mask coronagraph (PIAACMC) concepts, previously described for circular unobstructed apertures in \cite{2010ApJS..190..220G}, are here adapted to arbitrarily shaped apertures. Section \ref{sec:APCMLC} describes how the APCMLC can be adapted to non-circular apertures, and a step by step process to design a APCMLC for any aperture shape is proposed and examples are shown. In Section \ref{sec:PIAACMC}, the PIAACMC is shown to offer performance superior to the APCMLC, and its design for centrally obscured and segmented apertures is discussed, with examples representative of current and future large telescopes shown. High performance APCMLC and PIAACMC for pupils with strong aspect ratios is briefly discussed in Section \ref{sec:aspectratio}. Chromaticity of the concepts is discussed in Section \ref{sec:chrom}. Results are discussed in Section \ref{sec:conclusion}.

\begin{figure*}
\includegraphics[scale=0.93]{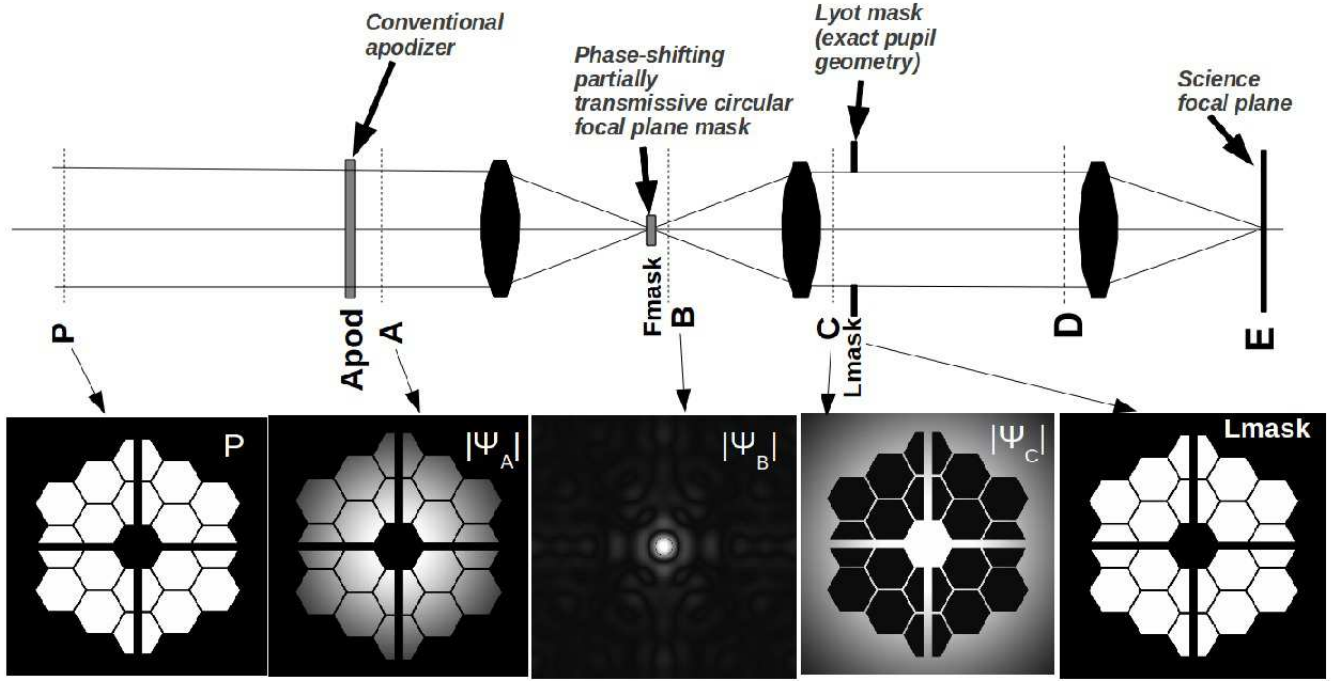} 
\caption{\label{fig:APCMLCprinciple} 
Apodized Pupil Complex Mask Lyot Coronagraph (APCMLC) design for a centrally obscured segmented aperture. The entrance aperture (complex amplitude P) is apodized (complex amplitude A) with a conventional apodizer. The central part of the corresponding on-axis PSF is both attenuated and phase-shifted (complex amplitude B) by the focal plane mask, yielding perfect destructive interference within the geometric pupil (as shown by pupil complex amplitude C). The Lyot mask (Lmask) rejects all light from the on-axis source, while it transmits all of the light from distant off-axis sources. 
}
\end{figure*}

\begin{figure*}
\includegraphics[scale=0.37]{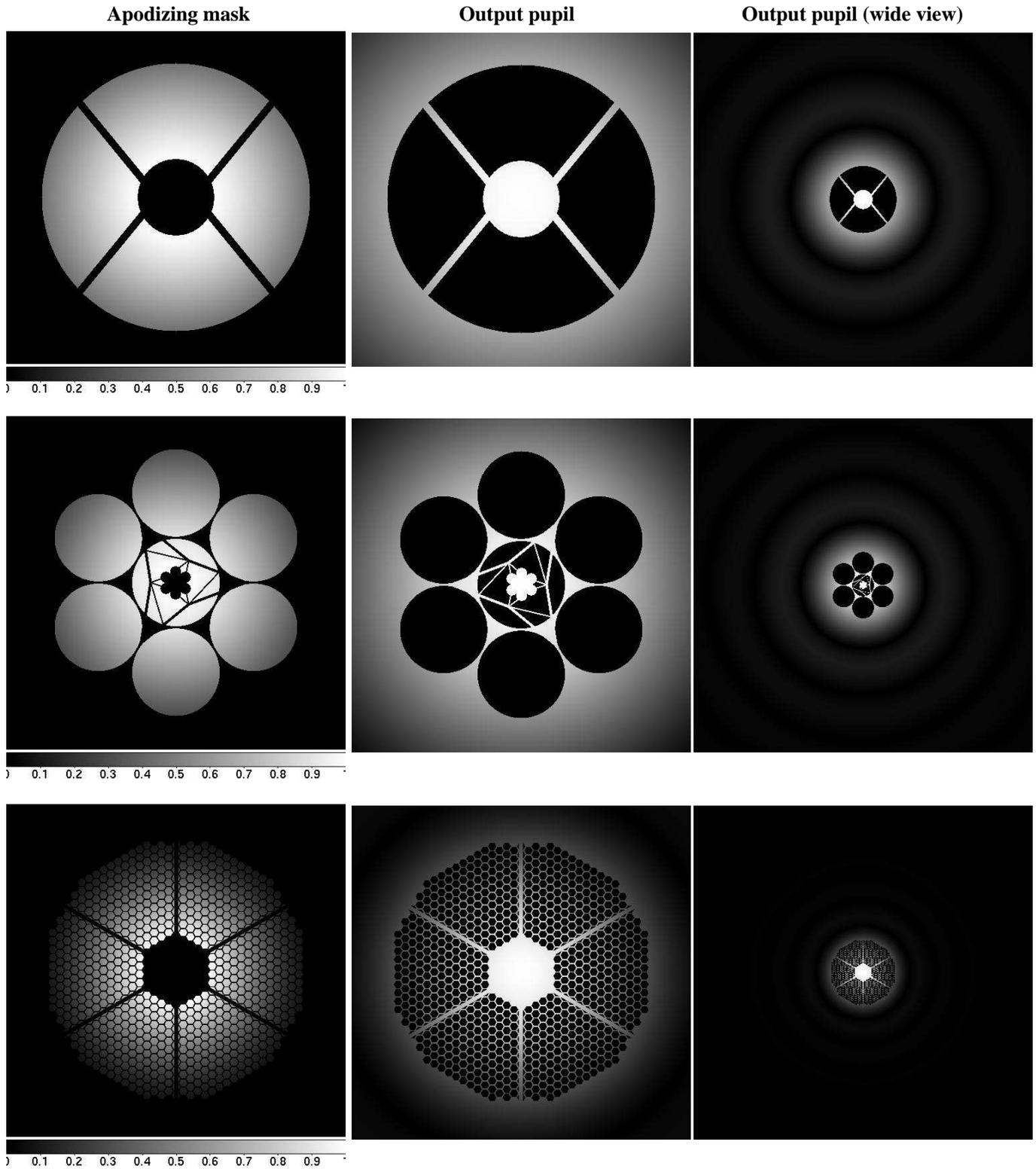} 
\caption{\label{fig:APCMLCexamples} 
Pupil plane intensity apodization function (left), and Lyot plane amplitude distribution for an on-axis point source (center and right) for several APCMLC designs. Top: APCMLC design \#1 for the Subaru Telescope pupil. Center: APCMLC design \#2 for the GMT design. Bottom: APCMLC design \#3 for the E-ELT pupil.
}
\end{figure*}

\begin{deluxetable*}{lcccc}
%\begin{deluxetable}{lcccc}  % for single column
\tabletypesize{\footnotesize}
\tablecolumns{5} 
\tablewidth{0pc} 
\tablecaption{\label{tab:APCMLC} APCMLC design examples for segmented apertures} 
\tablehead{ 
\colhead{Design} & \colhead{Focal plane mask radius (a/2)} & \colhead{Focal plane mask transm\tablenotemark{a}} & \colhead{Throughput\tablenotemark{b}} & \colhead{IWA\tablenotemark{c}} }

\startdata 
\multicolumn{5}{l}{{\bf Subaru Telescope pupil}}\\
Subaru APCMLC \#1   & 0.596 $\lambda/D$ & 99.62\% & 68.88\% & 0.71 $\lambda/D$\\
Subaru APCMLC \#2   & 0.8 $\lambda/D$ & 24.89\% & 54.65\% & 0.90 $\lambda/D$\\
Subaru APCMLC \#3   & 1.2 $\lambda/D$ & 8.57\% & 39.19\% & 1.30 $\lambda/D$\\
\multicolumn{5}{l}{{\bf Giant Magellan Telescope (GMT) pupil}}\\
GMT APCMLC \#1      & 0.666 $\lambda/D$ & 99.64\% & 64.50\% & 0.78 $\lambda/D$\\
GMT APCMLC \#2      & 0.7 $\lambda/D$ & 79.47\% & 61.99\% & 0.81 $\lambda/D$\\
GMT APCMLC \#3      & 1.2 $\lambda/D$ & 35.16\% & 11.39\% & 1.28 $\lambda/D$\\
GMT APCMLC \#4      & 1.5 $\lambda/D$ & 28.59\% & 9.68\% & 1.58 $\lambda/D$\\
\multicolumn{5}{l}{{\bf Thirty Meter Telescope (TMT) pupil}}\\
TMT APCMLC \#1      & 0.764 $\lambda/D$ & 99.72\% & 55.67\% & 0.86 $\lambda/D$\\
TMT APCMLC \#2      & 0.8 $\lambda/D$ & 85.48\% & 53.23\% & 0.90 $\lambda/D$\\
TMT APCMLC \#3      & 1.2 $\lambda/D$ & 36.08\% & 34.26\% & 1.27 $\lambda/D$\\
TMT APCMLC \#4      & 1.5 $\lambda/D$ & 40.94\% & 28.41\% & 1.57 $\lambda/D$\\
\multicolumn{5}{l}{{\bf European Extremely Large Telescope (E-ELT) pupil}}\\
E-ELT APCMLC \#1    & 0.825 $\lambda/D$ & 99.85\% & 54.26\% & 0.93 $\lambda/D$\\
E-ELT APCMLC \#2    & 0.9 $\lambda/D$ & 78.76\% & 49.92\% & 1.00 $\lambda/D$\\
E-ELT APCMLC \#3    & 1.2 $\lambda/D$ & 50.56\% & 38.43\% & 1.29 $\lambda/D$\\
\enddata 
\tablenotetext{a}{All focal plane masks are phase shifting (half wave) in addition to being partially transmissive. For each pupil, the first design is meant to approximate a fully transmissive phase shifting focal plane mask, but the transmission is not exactly 100\%, as the design was obtained by setting the focal plane mask radius to a multiple of 0.001 $\lambda$/D.}
\tablenotetext{b}{System throughput is equal to the overall intensity transmission of the pupil plane apodizer}
\tablenotetext{c}{Angular separation at which the throughput is 50\% of the pupil apodizer throughput}
\end{deluxetable*} 
%\end{deluxetable} % for single column

\section{Apodized pupil complex mask Lyot Coronagraph (APCMLC) for apertures of arbitrary shape}
\label{sec:APCMLC}

\subsection{Principle}
\label{ssec:APCMLCprinc}

In this section, it is shown that the APCMLC is compatible with non-circular apertures, as illustrated in Figure \ref{fig:APCMLCprinciple}, and a description of how it can be designed for arbitrarily shaped pupils is provided. While the APCMLC description provided here remains qualitative and focused on aspects relevant to non-circular apertures, a more complete analytical description is provided in \cite{2010ApJS..190..220G} for circular unobstructed apertures. We describe the APCMLC by following how electric field (also refered to as complex amplitude in this paper) from an on-axis point source propagates through the coronagraph system.

The APCMLC, illustrated in Figure \ref{fig:APCMLCprinciple}, uses a circular focal plane mask to partially transmit and phase shift the on-axis point spread function (PSF) core (complex amplitude B on Figure \ref{fig:APCMLCprinciple}). The transmission and phase shift are uniform within the mask radius, and the mask is fully transmissive, with no phase shift, outside this radius. This produces a destructive interference within the geometric pupil, between the light that passes around the focal plane mask disk and the phase-shifted light passing through the focal plane phase-shifting disk. With a Lyot mask (Lmask) selecting only the geometric pupil, a coronagraphic effect is achieved. The concept is thus an intermediate point between the conventional Lyot coronagraph or Apodized Pupil Lyot Coronagraph (APLC) \citep{2003AA...397.1161S}, which use a large fully opaque focal plane mask, and the phase mask coronagraph \citep{1997PASP..109..815R,2000SPIE.4006..377G,1999PASP..111.1321G,2010AA...509A...8N} which uses a small size fully transmissive phase-shifting focal plane mask. In the APCMLC, the focal plane mask size can be chosen anywhere between these two extremes, and defines the ratio between the amount of light within the circular mask and outside the mask. As the focal plane mask radius decreases, a smaller fraction of the light is within the mask radius, and its transmission must increase to maintain the flux balance between the "inside focal plane mask" (corresponding to low spatial frequencies in the pupil) and "outside focal plane mask" (high spatial frequencies in the pupil) components, a necessary condition to achieve destructive interference.

Full destructive interference within the geometric pupil also requires that the two components are equal in amplitude for every point within the pupil. Since this match does not naturally occur, all three concepts (APLC, Roddier phase mask coronagraph and APCMLC) require the entrance pupil to be amplitude apodized to enforce this match. Qualitatively, for small focal plane mask size, the apodization mostly changes the pupil light distribution for the ``outside focal plane mask'' component, while the light distribution for the ``inside focal plane mask'' component is mostly driven by the size of the focal plane mask. The entrance pupil apodization can therefore be iteratively derived to force the ``outside focal plane mask'' component to match the ``inside focal plane mask component'', using the following steps: 
\begin{enumerate}
\item{Adopt a focal plane mask diameter $a$}
\item{Compute the on-axis complex amplitude PSF for the aperture. This is the Fourier transform of the pupil complex amplitude P}
\item{Clip the PSF: values outside the focal plane mask radius are forced to zero}
\item{Inverse-Fourier transform the clipped PSF, and adopt this function as the apodized pupil plane amplitude function A, after multiplication by a factor $\Lambda_a$ so that its maximum value across the pupil is be equal to 1 (full transmission)}
\item{Return to step (2), with the output of step (4) as the pupil complex amplitude function}
\end{enumerate}

This iterative algorithm is a generalization of the iterative algorithm used by \cite{2000SPIE.4006..377G,2002AA...391..379G} and detailed in \citet{guyonPhD} to compute optimal apodization for the phase mask coronagraph (for which the mask is fully transmissive), and the iterative algorithm used to compute optimal apodization for the APLC (for which the mask is fully opaque) on centrally obscured circular apertures \citep{2005ApJ...618L.161S,2010AA...520A.110M} and on more complex pupil shapes \citep{2009ApJ...695..695S}. \citet{2002A&A...389..334A,2003AA...397.1161S} showed that the apodization solutions obtained for rectangular and circular apertures are Prolate functions for which analytical expressions exist. Apodization functions can also be computed for centrally obscured apertures \citep{2005ApJ...618L.161S}, and for arbitrary non circular symmetric pupils \citep{2009ApJ...695..695S}. A remarkable property of the iterative algorithm described above is that it converges for a wide range of pupil shapes and focal plane mask diameters \citep{guyonPhD}. For small focal plane mask diameters, convergence is due to the fact that modifying the entrance aperture light distribution predominantly affects the ``outside focal plane mask'' light component. Exact apodization solutions for the APLC and APCMLC therefore exist for most aperture geometries and focal plane mask diameters. An example APCMLC design on a non-circular aperture, for which the entrance pupil apodization function was computed using the iterative algorithm described in this section, is shown in Figure \ref{fig:APCMLCprinciple}.

The APCMLC is described here analytically for monochromatic light using notations shown in Figure \ref{fig:APCMLCprinciple}. The entrance pupil shape is defined by the real function $P(\mathbf{r})$, with $\mathbf{r}$ the 2-D position vector in the pupil plane, and $P(\mathbf{r})=1$ for points within the pupil and $P(\mathbf{r})=0$ outside of the pupil. The apodizer function $Apo(\mathbf{r})$ is applied to the pupil, yielding the following complex amplitude in plane $A$: 

\begin{equation}
\Psi_A(\mathbf{r}) = Apo(\mathbf{r}) P(\mathbf{r})
\end{equation}

The $\mathbf{r}$-dependence is dropped in subsequent equations. The iterative algorithm previously described is used to numerically compute the apodization function $Apo(\mathbf{r})$, which will converge to a pupil function $\psi_a$ which is the eigenvector of the "truncate (by P), Fourier Transform, truncate (by $|\mathbf{r}|<a$), and inverse Fourier Transform" operator, with eigenvalue equal to the scaling factor $\Lambda_a$ used in step (4) of the iterative algorithm given previously. 

\begin{equation}
\label{equ:La}
( \psi_a P ) \otimes \widetilde{M_a} = \Lambda_a \psi_a 
\end{equation}

 where $\otimes$ is the convolution operator, $M_a$ is is equal to 1 within a disk of diameter $a$ and is equal to 0 outside it, and $\widetilde{M_a}$ is the Fourier Transform of $M_a$. In the APCMLC, the apodizer function is chosen equal to $\psi_a$:

\begin{equation}
\Psi_A = \psi_a P
\end{equation}

The focal plane mask complex amplitude is :

\begin{equation}
F_{mask} = 1 - (1-t) M_a
\end{equation}

where $t$ is the complex amplitude transmission within the circular focal plane mask. The complex amplitude in plane B is : 

\begin{equation}
\Psi_B = F_{mask} \widetilde{\Psi_A} = \widetilde{\Psi_A} - (1-t) M_a \widetilde{\Psi_A}
\end{equation}

The complex amplitude in plane C is obtained by Fourier transform of $\Psi_B$:
\begin{equation}
\label{equ:PsiC0}
\Psi_C = \Psi_A - (1-t) (\psi_a P) \otimes \widetilde{M_a}
\end{equation}

By combining equations \ref{equ:La} and \ref{equ:PsiC0}, and multiplying by $P(r)$, the complex amplitude in plane C within the geometrical pupil is: 

\begin{equation}
\label{equ:PsiC}
\Psi_C P(\mathbf{r}) = P(\mathbf{r}) \times (1- (1-t) \Lambda_a) \psi_a
\end{equation}

This equation shows that, if $t = 1-\Lambda_a^{-1}$ (this value is now noted $t_a$), then $\Psi_C$ is equal to zero within the geometric pupil. This is the condition for a APCMLC, which completely removes light from an on-axis point source, provided that a Lyot pupil plane mask $L_{mask}(\mathbf{r})=P(\mathbf{r})$ is used to only select light within the geometric pupil. Since $\Lambda_a < 1$, $t_a$ is negative: the focal plane mask is both partially transmissive and π-phase shifting. A coronagraphic solution requires $t_a>-1$, and therefore exists only if $\Lambda_a > 0.5$: the focal plane size needs to be sufficiently large so that light going through the mask can be balanced with light going outside the mask.

The same pupil apodization technique is used in the Apodized Pupil Lyot Coronagraph (APLC) to optimize the pupil entrance complex amplitude to the hard edged opaque focal plane mask \citep{2003AA...397.1161S}. In the APLC, $t=0$ in equation \ref{equ:PsiC}, and the coronagraphic extinction is therefore not total for an on-axis point source. Equation \ref{equ:PsiC} shows that the on-axis PSF in the final focal plane mask is an exact copy of the non-coronagraphic PSF, scaled by $(1-\Lambda_a)^2$ in intensity. For large focal plane masks diameter $a$, $\Lambda_a$ is close to 1, and the coronagraphic extinction is satisfactory. The APLC concept has been adopted for the Palomar Observatory high contrast imaging program \citep{2011PASP..123...74H} and the Gemini Planet Imager \citep{2008SPIE.7015E..31M} and has been validated in laboratory demonstrations \citep{2011AJ....142..119T}. The APCMLC concept is very similar to the APLC, the only fundamental difference being that its focal plane mask transmission is allowed to be non-zero, therefore allowing full coronagraphic extinction for any focal plane mask size $a$ for which $\Lambda_a > 0.5$.

\subsection{APCMLC designs for segmented apertures}
\label{ssec:APCMLCdesigns}

Apodized pupil complex mask Lyot coronagraphs (APCMLCs) were designed for the Subaru Telescope, Giant Magellan Telescope (GMT), Thirty Meter Telescope (TMT) and European Extremely Large Telescope (E-ELT) pupil geometries, following the process described in the previous section. For each pupil geometry, several focal plane mask sizes were chosen. The designs with the smallest possible focal plane mask sizes use full transmission $\pi$-phase shifting circular focal plane masks, and are referred to as optimal IWA APCMLC designs in this paper. As the focal plane mask size increases, it also becomes more opaque, the system throughput (which is equal to the apodizer throughput) decreases and the IWA increases. Results are summarized in Table \ref{tab:APCMLC}, and show that optimal IWA designs offer IWAs around 0.9 $\lambda/D$ and throughputs around 60\%. For all designs, the IWA is approximately equal to the focal plane mask radius, and the throughput decreases rapidly with increasing focal plane mask size: with a 1.2 $\lambda/D$ radius, the throughput ranges from approximately 10\% to 35\% depending on the pupil geometry. The performance of the optimal IWA design is largely independent of pupil geometry, and is similar for segmented apertures to the performance previously reported for a full unobstructed circular pupil \citep{2010ApJS..190..220G}. However, as the mask size increases, pupil geometry has a larger impact on performance, as the range of pupil plane spatial frequencies accessed by the focal plane mask begins to overlap with the low spatial frequency components of the pupil geometry (central obstruction, large segments, thick spider vanes). This difference is most noticeable between the GMT pupil with few large segments and the TMT or EELT geometries with numerous small segments.

Selected examples of apodization functions and Lyot plane intensity images are shown in Figure \ref{fig:APCMLCexamples}. In each case, the apodization function is smooth and free of high spatial frequencies, and no light is left within the geometric pupil in the Lyot pupil plane, as all residual starlight is diffracted outside of the pupil and in the gaps between segments.

\begin{figure*}[tb]
\includegraphics[scale=1.02]{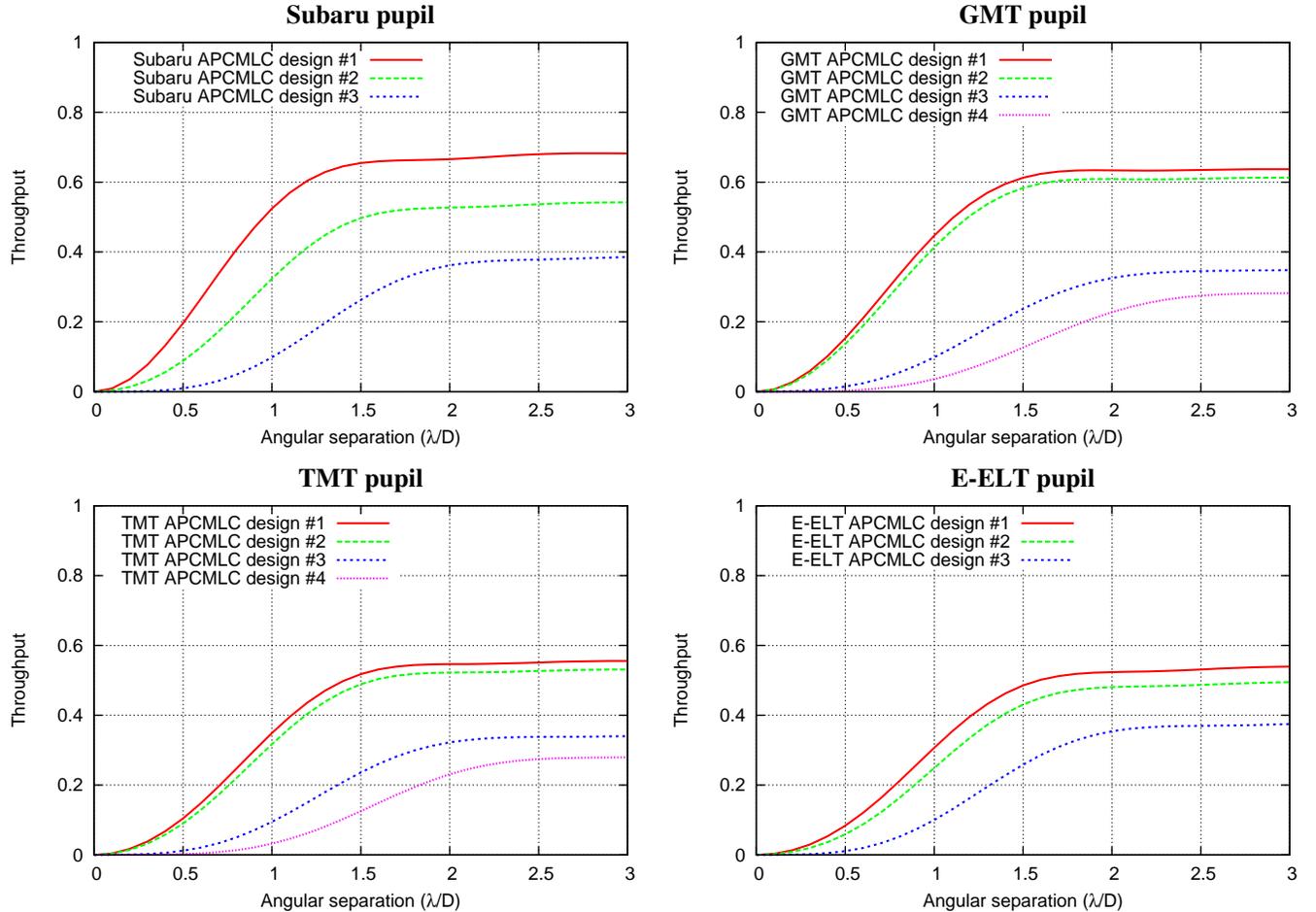} 
\caption{\label{fig:APCMLCtransm} 
Coronagraph system transmission for a monochromatic point source, as a function of the angular separation between the point source and the optical axis, for the APCMLC designs listed in Table \ref{tab:APCMLC}. System transmission is measured by integrating all light in the final focal plane (plane E in figure \ref{fig:APCMLCprinciple}).}
\end{figure*}

Table \ref{tab:APCMLC} gives for several APCMLC designs the key design parameters (focal plane mask size $a$, focal plane mask transmission) as well as the coronagraph performance (throughput and IWA). For each pupil shape considered, the first design (design \# 1) is the most aggressive in IWA, with a nearly fully transmissive focal plane mask. This aggressive design is also the one with the highest throughput, as the apodization strength needs to increase for larger focal plane mask sizes. The APCMLC throughput never exceeds 70\% due to the need for a pupil apodization. Transmission curves are given in Figure \ref{fig:APCMLCtransm} for the APCMLC designs listed in Table \ref{tab:APCMLC}.

\section{Phase Induced Amplitude Apodization Complex Mask Coronagraph (PIAACMC) for apertures of arbitrary shape}
\label{sec:PIAACMC}

\subsection{Lossless Phase-Induced Amplitude Apodization (PIAA)}

\begin{figure*}[tb]
\includegraphics[scale=0.93]{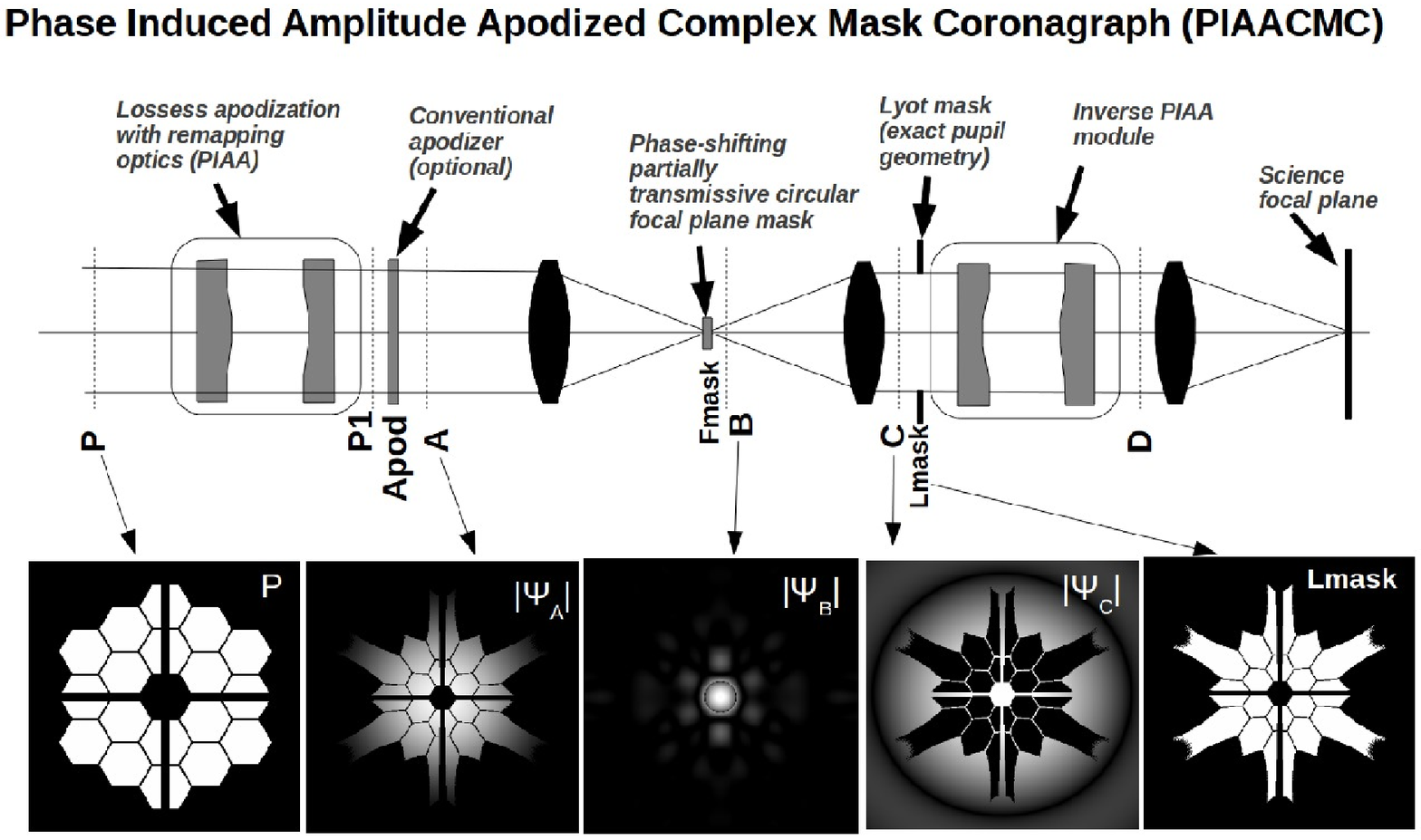} 
\caption{\label{fig:PIAACMCprinc} 
PIAACMC design for a centrally obscured segmented aperture. The entrance aperture (P) is apodized (P1) thanks to aspheric PIAA optics. The central part of the corresponding on-axis PSF is both attenuated and phase-shifted (B) by the focal plane mask, yielding perfect destructive interference within the geometric pupil (C). The Lyot mask (Lmask) rejects all light from the on-axis source, while it transmits all of the light from distant off-axis sources. Inverse PIAA optics can be introduced to recover a sharp off-axis image over a wide field of view. 
}
\end{figure*}

While the APCMLC described in Section \ref{sec:APCMLC} achieves full on-axis coronagraphic extinction for almost any pupil shape, its throughput is limited by the entrance apodization required to achieve perfect destructive interference in the output pupil plane. The system throughput decreases as the focal plane mask size increases, with a maximum throughput equal to 72\% for a 0.64 $\lambda/D$ radius purely phase-shifting transparent mask on a circular unobstructed aperture. Throughput, and consequently angular resolution, degrade rapidly with increased focal plane mask size: it is 18\% for a 2 $\lambda/D$ radius mask. The results obtained in Section \ref{ssec:APCMLCdesigns} also show that the APCMLC maximum throughput (achieved for the designs with the most aggressive IWA) is lower on segmented pupils than it is for an unobstructed circular pupil ("Throughput" column of Table \ref{tab:APCMLC}). Moreover, throughput, angular resolution and IWA are significantly degraded when the focal plane mask size is increased - while mitigation of undesired chromatic effects at the focal plane mask may require a larger and more opaque mask.

Phase-induced Amplitude Apodization (PIAA) uses aspheric mirrors to achieve a lossless beam apodization \citep{2003AA...404..379G}, and can therefore produce a highly apodized beam suitable for high contrast imaging without the angular resolution loss and throughput loss of a conventional apodizer. PIAA can also be used to replace the entrance apodization in the APCMLC design described in Section \ref{sec:APCMLC}, as previously proposed for unobstructed circular pupils \citep{2010ApJS..190..220G}. The resulting coronagraph, denoted Phase-induced Amplitude Apodization Complex mask coronagraph (PIAACMC), offers simultaneously full throughput, small inner working angle and total on-axis extinction.

An example PIAACMC design is shown in Figure \ref{fig:PIAACMCprinc} for a segmented centrally obscured pupil. The entrance pupil P (image shown in the lower left of the figure) is apodized with lossless aspheric PIAA optics. Because the PIAA optics perform apodization by remapping instead of selective transmission, the resulting pupil P1 shape is modified. A conventional apodizing mask may be used to fine-tune the apodization if the PIAA optics do not exactly produce the required amplitude distribution (this will be addressed in the following section). The resulting pupil A is shown in the second image from the lower left corner. The image of an on-axis point source is shown in the center image, where the phase-shifting partially transmissive focal plane mask is inserted. In the output pupil plane C, all light within the pupil has been removed, while diffracted starlight fills the gap and obstructions of the segmented pupil. A Lyot mask (noted Lmask) can then select only the geometric pupil area (after remapping) to fully block on-axis starlight while fully transmitting the light from distant off-axis source. A well-documented side-effect of apodization with PIAA optics is that off-axis PSFs are highly distorted, and corrective optics (inverse PIAA) are required at the output of the coronagraph to maintain diffraction limited sharp PSFs over a scientifically useful field of view \citep{2009PASP..121.1232L}. Except for PIAA and inverse PIAA optics, the PIAACMC architecture is functionally identical to the APCMLC architecture described in Section 2.1: between planes P1 (output of the PIAA optics) and the plane immediately after the pupil plane Lyot mask, the architecture is an APCMLC. The main difference between APCMLC and PIAACMC is that the lossless apodization allows increased performance by maintaining full throughput and angular resolution, regardless of the focal plane mask size adopted.

\subsection{Designing a PIAACMC for a non circular aperture}

We consider in this work PIAACMC designs that perform a lossless PIAA apodization of the pupil to produce a generalized prolate function for the aperture. We  note that other apodization functions could be adopted, and could potentially lead to superior performance, but this is not explored in this paper. In the unobstructed circular pupil case \citep{2010ApJS..190..220G}, designing the PIAACMC is relatively simple, as PIAA apodization using a radial remapping function preserves the circular aperture shape. The prolate function can thus be first computed, and then realized with a radial PIAA apodization.

Designing a PIAACMC for complex shaped apertures is considerably more challenging because the PIAA apodization modifies the aperture shape, which itself changes the generalized prolate function. In addition to this circular problem, if the aperture is not circularly symetric, the generalized prolate is also not symmetric, and the required remapping function therefore cannot be written as a radial function. While PIAA optics can be designed for any radial remapping \citep{2003AA...404..379G}, an arbitrarily chosen 2D remapping function can almost never be realized with a set of two PIAA optics.

To overcome the two challenges listed above (circular design problem due to effect of PIAA on aperture shape, and complexity/impossibility of designing PIAA optics for non-circular symmetric remapping), a hybrid PIAACMC design is adopted, which includes a conventional apodizer after the remapping to produce the required prolate function. Thanks to this post-apodizer, the output of the PIAA apodization does not need to exactly match the generalized prolate function, allowing radial remapping functions to be used on non-circular symmetric apertures. The goal of the design optimization is to bring the PIAA apodization and generalized prolate functions close, in order to minimize the strength of the post-apodizer and thus maintain a high system throughput.

The proposed steps for designing a PIAACMC for complex shaped apertures are :
\begin{enumerate}
\item{Choose radial remapping function $r_1=f_b(r_0)$, where $r_0$ and $r_1$ are the radial coordinate in the input (before remapping) and output(after remapping) pupils respectively. For convenience, the remapping function is selected among a pre-computed set of functions used to produce prolate spheroidal apodizations on circular apertures. The focal plane mask diameter corresponding to the prolate spheroidal function is denoted $b$, and the corresponding remapping function and apodization intensity profile are respectively $f_b$ and $I_b$.}
\item{Apply the remapping function to the entrance pupil. The remapping transforms the entrance pupil intensity P(x,y) into P1(x,y).}
\item{Choose a focal plane mask diameter $a$.}
\item{Compute the generalized prolate function $Prola(x,y)$ for the remapped aperture shape defined by $P1(x,y)>0$, using the focal plane mask diameter $a$. This is done iteratively as described in section \ref{ssec:APCMLCprinc}}
\item{Compute the amplitude ratio $Apo(x,y)=Prola(x,y)/P1(x,y)$. This is the post-apodizer amplitude transmission function. $Apo(x,y)$ is then scaled to ensure that its maximum value is equal to 1. The intensity-weighted average of $Apo(x,y)^2$ defines the coronagraph throughput for off-axis sources.}
\end{enumerate}

Steps (3) to (5) are repeated for different values of the focal plane mask size $a$. The off-axis coronagraph throughput is computed for each choice of $a$, and the final focal plane mask size is chosen to maximize throughput. This optimization links the choice of the remapping function (step (1)) to a value of the focal plane mask radius a. For a circular unobstructed aperture, the solution would be $a=b$, for which the PIAA apodization would perfectly match the generalized prolate function. On arbitrarily shaped pupils, the focal plane mask radius is usually close to, but not equal to, $b$. Stronger apodization functions correspond to larger values for $a$ and $b$.

\begin{figure*}[tb]
\includegraphics[scale=0.3]{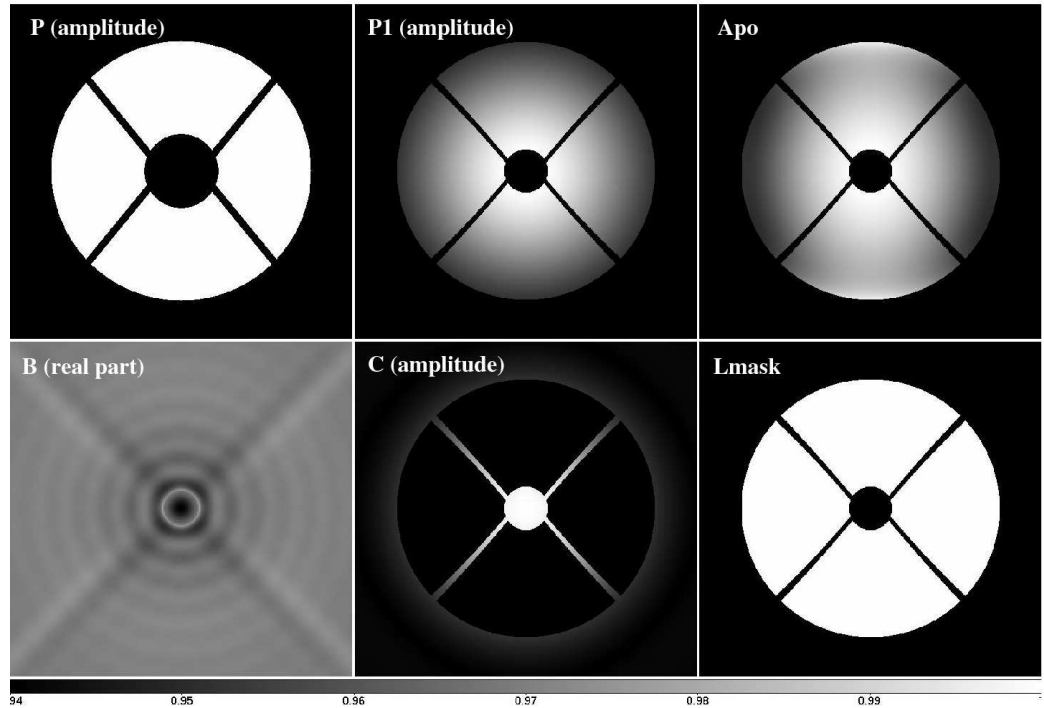} 
\caption{\label{fig:PIAACMC_Subaru} 
Two PIAACMC designs for the Subaru Telescope pupil: small IWA design (left) and large IWA design (right). The entrance pupil P (top left) is remapped into P1 (top center) with PIAA optics. A high transmission apodizer (top right) slightly modifies P1 into the desired generalized Prolate $\psi_a$ for the corresponding aperture shape. A circular phase-shifting partially transmissive mask is introduced in the focal plane, producing the complex amplitude B (bottom left shows real part of B) which is a real function). The corresponding pupil plane complex amplitude (bottom center) shows total destructive interference within the pupil. The brightness scale at the bottom of the figure applies to the apodizing mask in the top right. 
}
\end{figure*}

\begin{figure}[tb]
\includegraphics[scale=0.42]{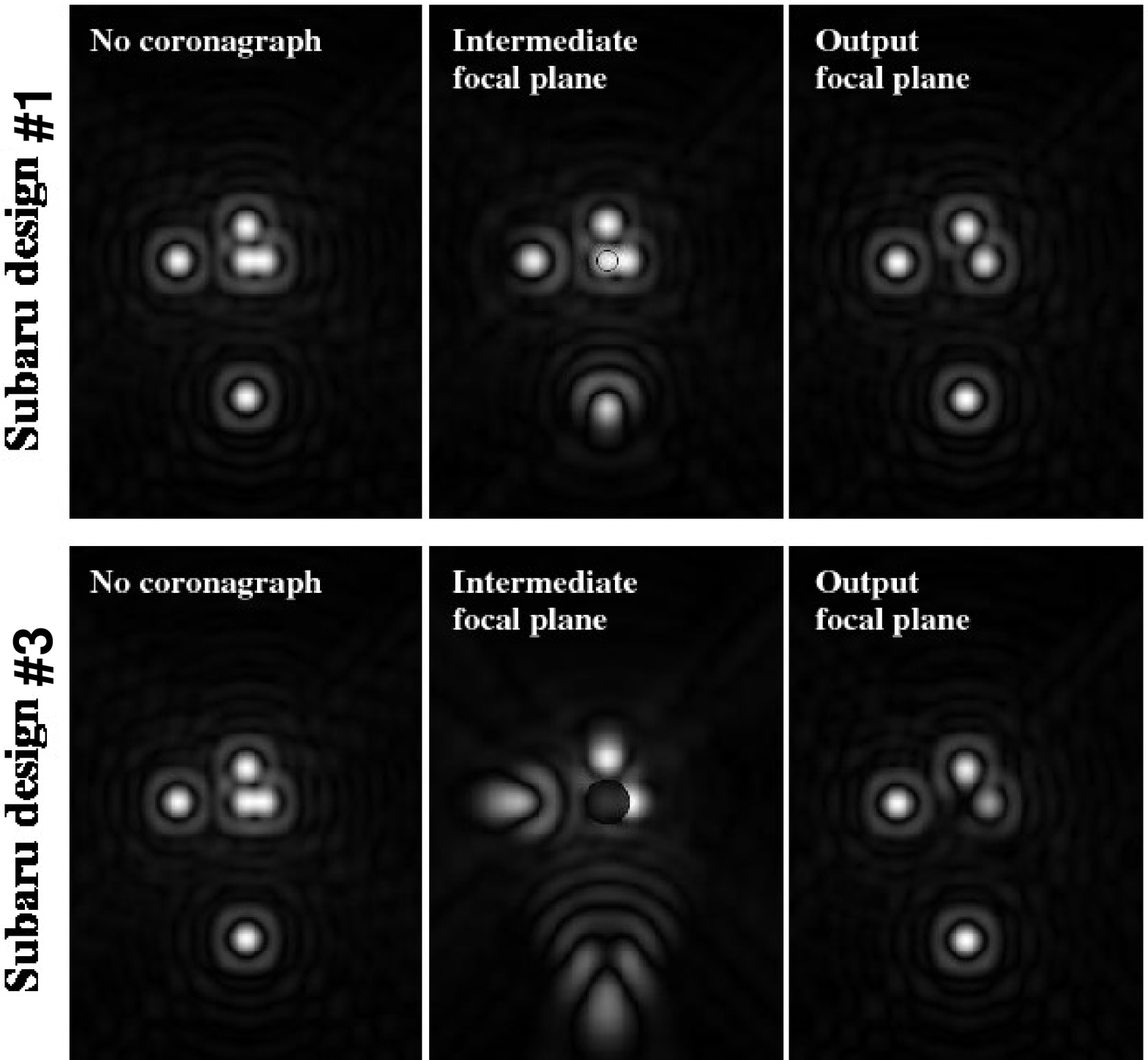} 
\caption{\label{fig:PIAACMC_Subaru_PSFs} 
Simulated Subaru PIAACMC images of 5 point sources of equal brightness. The point sources are at coordinates (0;0), (1;0), (0;2), (-4;0) and (0;-8) in $/lambda/D$ units. A non-coronagraphic images (left) shows all five point sources. The partially transmissive central focal plane mask is visible in the intermediate focal plane image (center), where off-axis PSFs are distorted by the PIAA remapping. In the output focal plane image (right), the central source is fully canceled and the off-axis PSFs images are sharp thanks to the inverse PIAA optics.
}
\end{figure}

\subsection{PIAACMC design examples}

\subsubsection{Centrally obscured pupils: Subaru Telescope pupil}

The Subaru telescope pupil is representative of current large aperture astronomical telescopes, with a large central obstruction and thick spiders. Both features must be taken into account for the design of a high performance coronagraph.

Figure \ref{fig:PIAACMC_Subaru} shows two PIAACMC designs for the Subaru Telescope pupil. The small IWA design (left) was computed from the $b/2=0.6 \lambda/D$ beam remapping, and uses a small sub-$\lambda/D$ radius focal plane mask with high transmission. The large IWA design was computed from $b/2=1.2 \lambda/D$, adopts a larger mostly opaque focal plane mask, and relies on a stronger PIAA remapping. Both designs offer throughput above 97\%, and their throughput could be further increased by slightly elongating the focal plane mask, which was kept circular for simplicity in this study. The large IWA design, by relying on a stronger PIAA remapping, introduces a large pupil deformation, as visible in the figure. The post-focal plane mask pupil images demonstrate the PIAACMC's ability to diffract all of the light from a central source outside the geometrical pupil, including within the gaps of the pupil (here, central obstruction and spiders). 

Figure \ref{fig:PIAACMC_Subaru_PSFs} shows intensity images of a field consisting of five equally bright point sources. The left images are obtained without a coronagraph, and simply show the imaging quality of the Subaru pupil in the absence of wavefront aberrations. The center column shows images in plane B of Figure \ref{fig:PIAACMCprinc}, immediately after the focal plane mask. The focal plane mask in the low IWA design (top) is more transmissive, and is also physically smaller. The large IWA design (bottom) introduces large off-axis aberrations due to the stong remapping. In the final coronagraphic images (right column), the central source is perfectly removed, and the images of the off-axis sources are sharp thanks to the inverse-PIAA optics.

\subsubsection{Segmented pupils: Giant Magellan Telescope (GMT)}

The Giant Magellan Telescope (GMT) consists of one central 8.4-m circular segment, surrounded by a ring of six 8.4-m diameter segments. While the outer segments are unobscured, the central segment includes a central obstruction due to the secondary mirrors and its support structure.

\begin{figure*}[p]
% single column version
\includegraphics[scale=0.15]{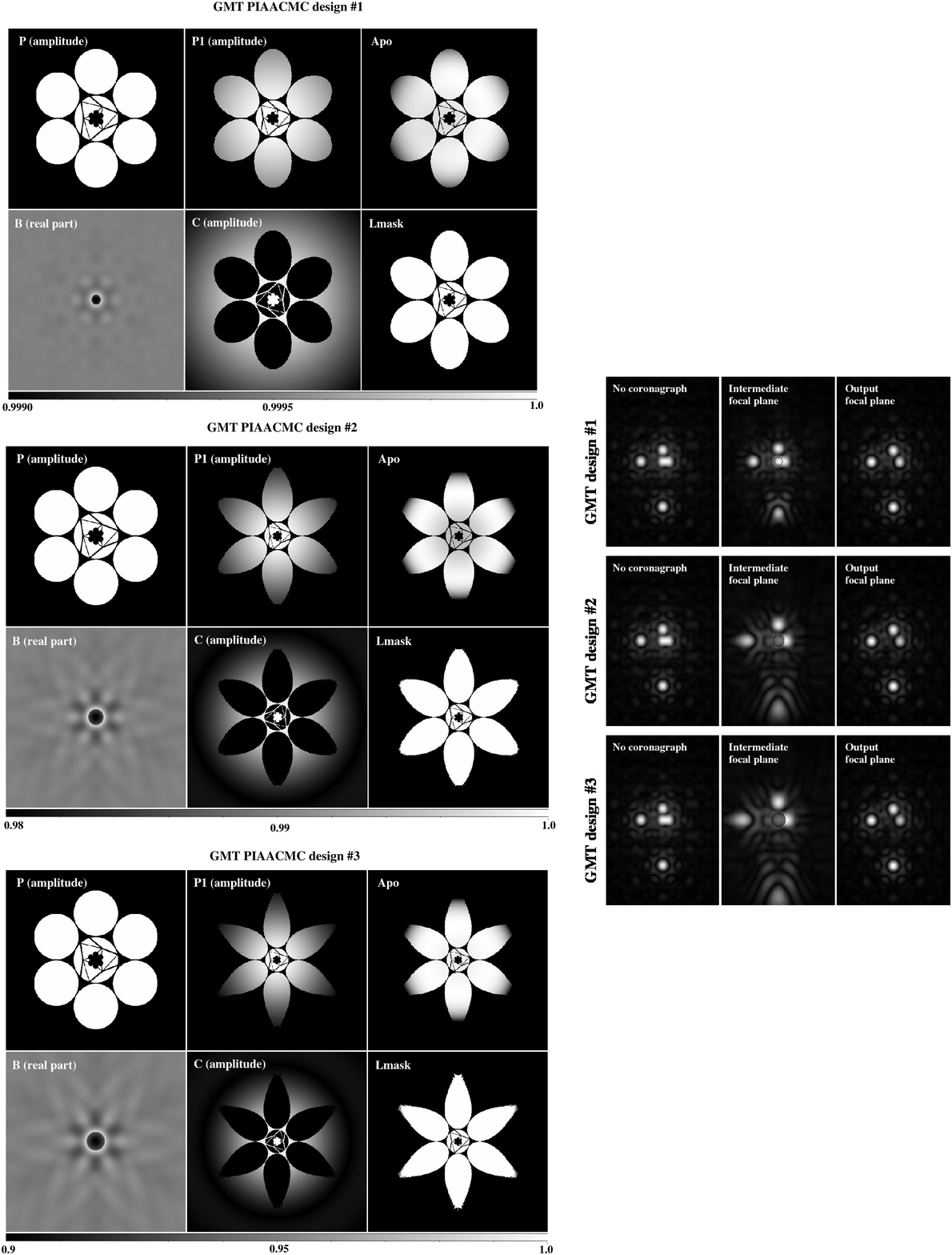} 
% double column version
%\includegraphics[scale=0.16]{fig07.eps} 
\caption{\label{fig:PIAACMC_GMT} 
Small (top left), medium (center left) and large (bottom left) IWA PIAACMC designs for the GMT pupil
}
\end{figure*}

Figure \ref{fig:PIAACMC_GMT} show three PIAACMC designs for the GMT pupil: a small IWA design computed for $b/2=0.7 \lambda/D$ (design \#1), a medium IWA design computed for $b/2=1.2 \lambda/D$ (design \#2), and a large IWA design computed for $b/2=1.5 \lambda/D$ (design \#3). As $b$ increases, the PIAA remapping becomes stronger, and the physical size of the focal plane mask increases. In each case, the PIAACMC achieves complete suppression of the on-axis point source, and its light is diffracted outside the geometrical aperture in plane C, including between the seven subapertures and within the secondary mirror obstruction and support structure. 

The imaging quality of the GMT PIAACMC designs is illustrated in the right panel of Figure \ref{fig:PIAACMC_GMT}, which shows that for each of the three designs, the final coronagraphic image maintains high thoughput and largely uncompromized imaging quality outside the central $\approx 1 \lambda/D$ region. The images also show that off-axis aberrations are stronger as the design relies more on PIAA remapping, although these aberrations are well corrected by the inverse PIAA system.

\subsubsection{Highly segmented pupils: European Extremely Large Telescope (E-ELT) and Thirty Magellan Telescope (TMT)}

\begin{figure*}[p]
\includegraphics[scale=0.15]{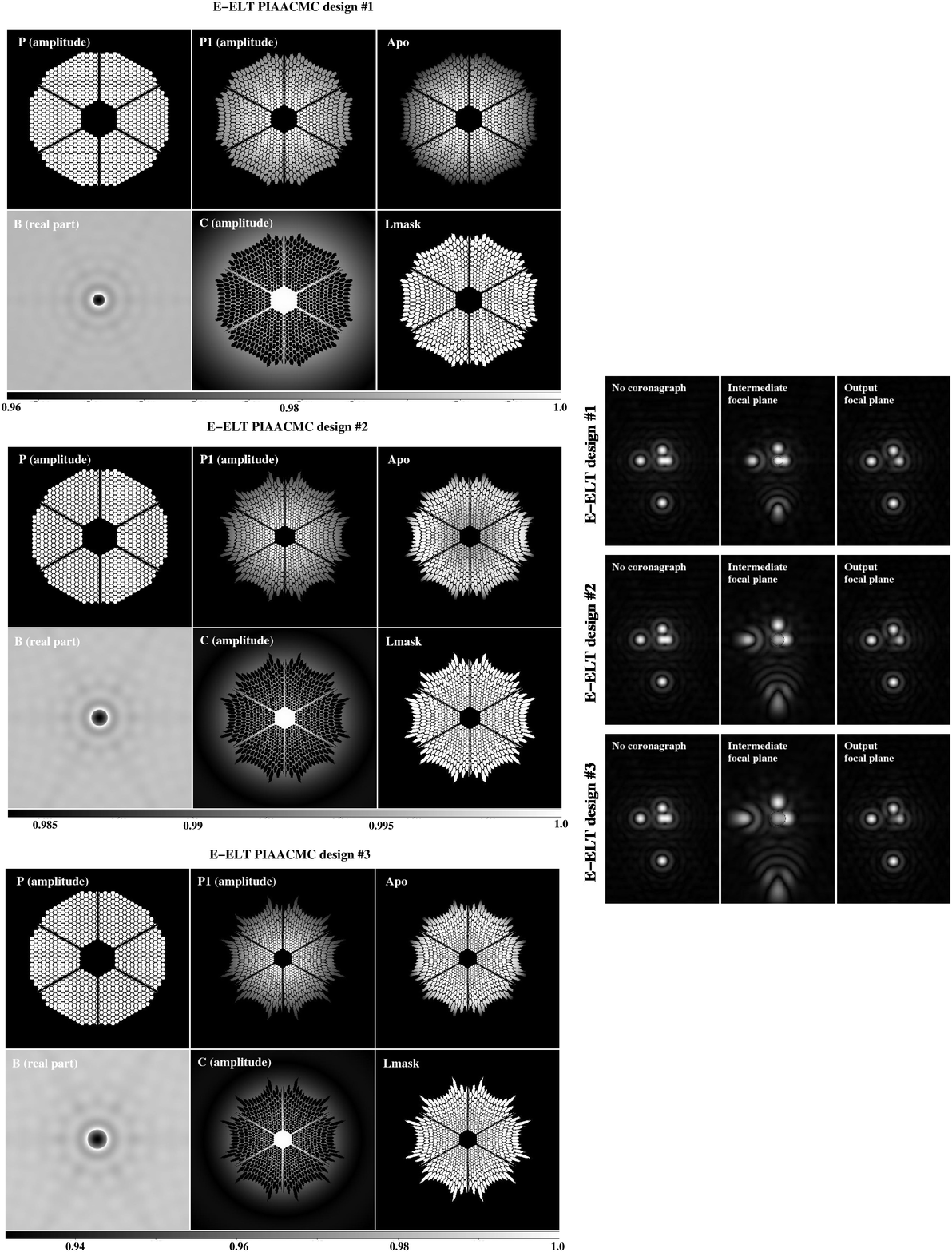} 
\caption{\label{fig:PIAACMC_EELT} 
Small (top left), medium (center left) and large (bottom left) IWA PIAACMC designs for the E-ELT pupil.
}
\end{figure*}

\begin{figure*}[p]
\includegraphics[scale=0.15]{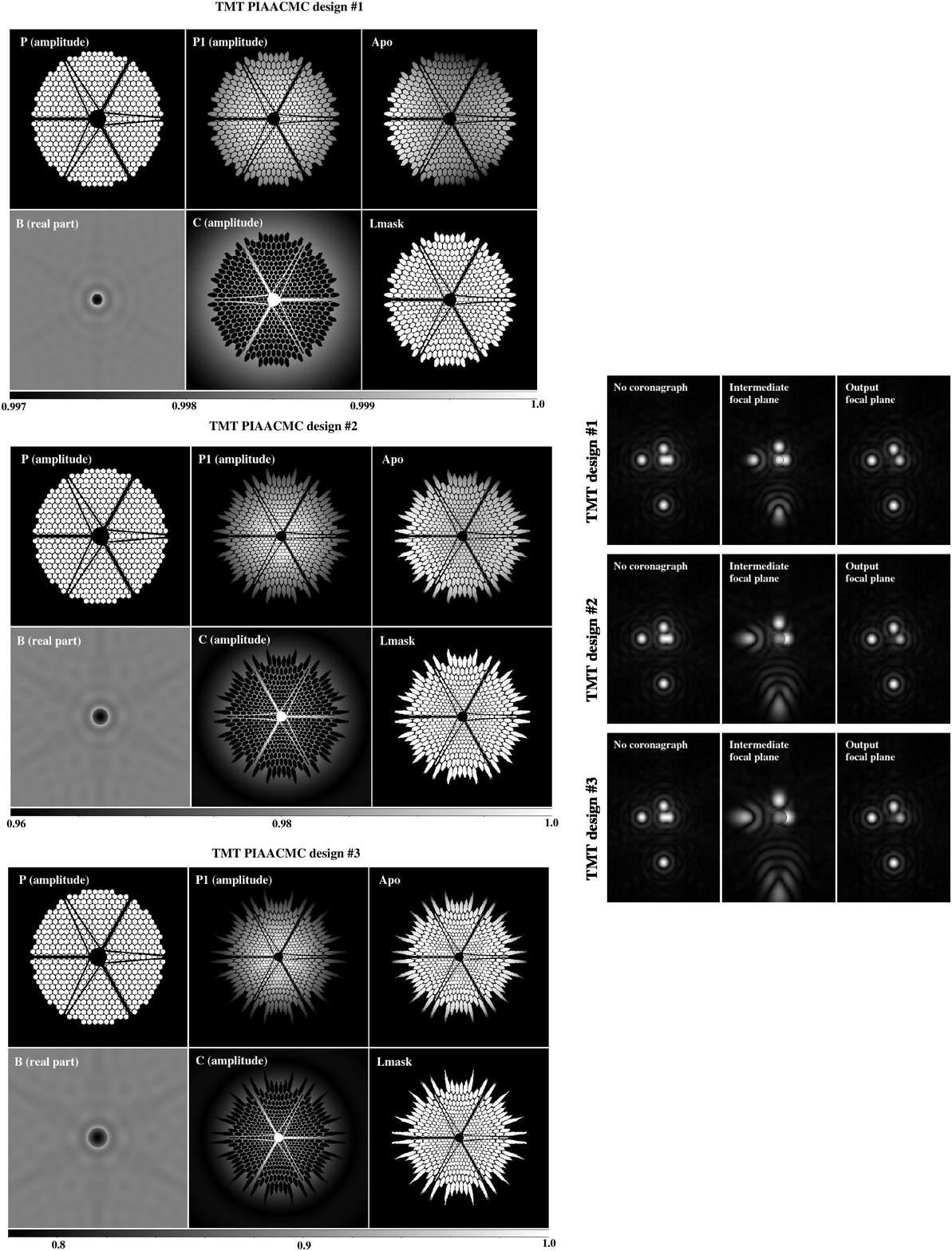} 
\caption{\label{fig:PIAACMC_TMT} 
Small (top left), medium (center left) and large (bottom left) IWA PIAACMC designs for the TMT pupil
}
\end{figure*}

Figures \ref{fig:PIAACMC_EELT} and \ref{fig:PIAACMC_TMT} each show three PIAACMC designs for the European Extremely Large Telescope (E-ELT) and the Thirty Meter Telescope (TMT) pupil geometries. Both pupils consist of a large number of small segments, a central obstruction and spider vanes. Each of the six designs achieves total rejection of a central point souce with high system throughput (between 97.8 \% and 99.98\%) for off-axis sources. The inner working angle ranges from $\approx$ 0.8 $\lambda$/D for the most aggressive designs (designs \#1) to $\approx$ 1.0 $\lambda$/D for the more conservative designs (designs \#3). Figures \ref{fig:PIAACMC_EELT} and \ref{fig:PIAACMC_TMT} show that thanks to the phase-shifting focal plane mask, light from an on-axis source is diffracted between the small segments of the pupil, within the spider vane shadows, within the central obstruction and outside the overall pupil: in the output pupil plane, no light is present within the geometric pupil.

For these designs, the Lyot mask must mask the gaps between the segements while transmitting light within the segments, and it must therefore be carefully aligned with the pupil. A Lyot mask for which the masked zones are slightly oversized may be used to accomodate pupil alignment errors at the cost of system throughput.

\subsection{Discussion}

\begin{deluxetable*}{lcccccc}
%\begin{deluxetable}{lcccccc}
% for single column
\tabletypesize{\footnotesize}
\tablecolumns{7} 
\tablewidth{0pc} 
\tablecaption{\label{tab:PIAACMC} PIAACMC design examples for segmented apertures} 
\tablehead{ 
\colhead{Design} & \colhead{b/2\tablenotemark{a}} & \colhead{PIAA strength\tablenotemark{b}} & \colhead{FPM rad\tablenotemark{c}} & \colhead{FPM transm} & \colhead{Throughput\tablenotemark{d}} & \colhead{IWA\tablenotemark{e}} }
%\colhead{      } & \colhead{(b/2)}      & \colhead{(Imax/Imin)}   & \colhead{radius (a/2)} & \colhead{mask transm} & \colhead{} & \colhead{} 

\startdata 
\multicolumn{7}{l}{{\bf Subaru Telescope pupil}}\\
Subaru PIAACMC \#1   & 0.6 $\lambda/D$   &  2.42   &  0.603 $\lambda/D_{syst}$  & 84.24\% & 99.91\% & 0.67 $\lambda/D$\\
Subaru PIAACMC \#2   & 0.9 $\lambda/D$   &  6.79   &  0.916 $\lambda/D_{syst}$  &  8.57\% & 99.39\% & 0.88 $\lambda/D$\\
Subaru PIAACMC \#3   & 1.2 $\lambda/D$   & 26.83   &  1.33  $\lambda/D_{syst}$  &  2.06\% & 97.04\% & 1.11 $\lambda/D$\\
\multicolumn{7}{l}{{\bf Giant Magellan Telescope (GMT) pupil}}\\
GMT APCMLC \#1      & 0.7 $\lambda/D$   &   3.30   &  0.693 $\lambda/D_{syst}$  & 98.55\% & 99.98\% & 0.72 $\lambda/D$\\
GMT APCMLC \#2      & 1.2 $\lambda/D$   &  26.83   &  1.12  $\lambda/D_{syst}$  & 20.71\% & 99.47\% & 0.89 $\lambda/D$\\
GMT APCMLC \#3      & 1.5 $\lambda/D$   & 124.09   &  1.32  $\lambda/D_{syst}$  & 16.64\% & 99.14\% & 0.92 $\lambda/D$\\
\multicolumn{7}{l}{{\bf Thirty Meter Telescope (TMT) pupil}}\\
TMT APCMLC \#1      & 0.8 $\lambda/D$   &   4.69   &  0.797 $\lambda/D_{syst}$  & 85.51\% & 99.80\% & 0.78 $\lambda/D$\\
TMT APCMLC \#2      & 1.2 $\lambda/D$   &  26.83   &  1.16  $\lambda/D_{syst}$  & 32.46\% & 98.51\% & 0.94 $\lambda/D$\\
TMT APCMLC \#3      & 1.5 $\lambda/D$   & 124.09   &  1.394 $\lambda/D_{syst}$  & 27.73\% & 98.71\% & 0.99 $\lambda/D$\\
\multicolumn{7}{l}{{\bf European Extremely Large Telescope (E-ELT) pupil}}\\
E-ELT APCMLC \#1    & 0.8 $\lambda/D$   &   4.69   &  0.816 $\lambda/D_{syst}$  & 99.87\% & 97.77\% & 0.81 $\lambda/D$\\
E-ELT APCMLC \#2    & 1.2 $\lambda/D$   &  26.83   &  1.15  $\lambda/D_{syst}$  & 45.58\% & 99.50\% & 0.93 $\lambda/D$\\
E-ELT APCMLC \#3    & 1.5 $\lambda/D$   & 124.09   &  1.37  $\lambda/D_{syst}$  & 37.85\% & 99.42\% & 0.98 $\lambda/D$\\
\enddata 
\tablenotetext{a}{The parameter b defines the pupil apodization function used for the PIAACMC design}
\tablenotetext{b}{PIAA apodization strength defined here as the ratio between surface brightness at the output beam center (Imax) and at the output beam edge (Imin)}
\tablenotetext{c}{Physical radius of the focal plane mask in units of $\lambda$/D for plane P1. Due to the slope amplification effect produced by remapping, this unit is different from angular coordinated on the sky.}
\tablenotetext{d}{Throughput values reflect small mismatch between the circular remapping adopted in this paper and the non-circular pupil geometry, rather than fundamental limitations of the PIAACMC concept.}
\tablenotetext{e}{Angular separation at which the throughput is 50\% of the pupil apodizer througput}
\end{deluxetable*} 
%\end{deluxetable} 

\begin{figure*}[tb]
\includegraphics[scale=1.0]{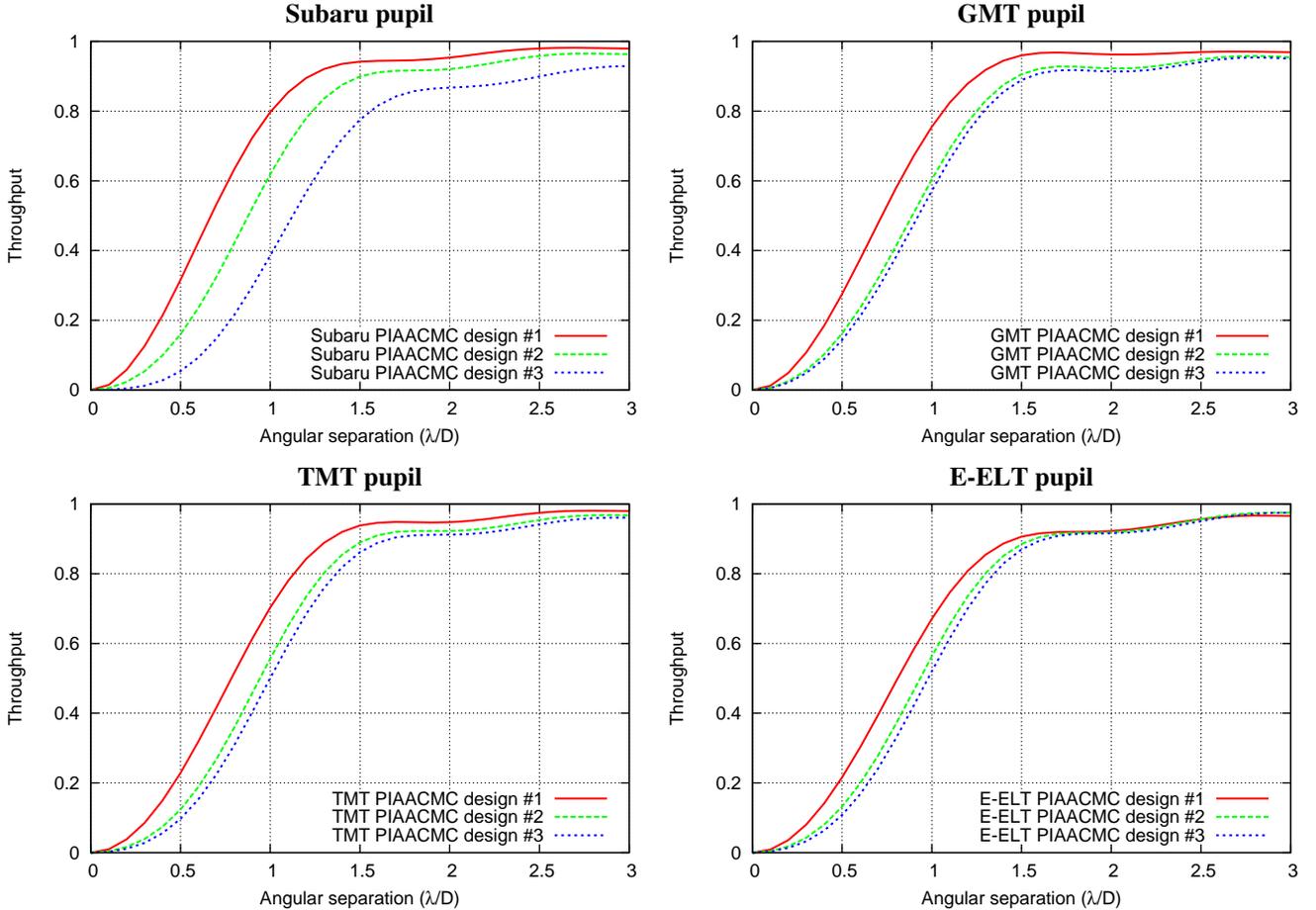} 
\caption{\label{fig:PIAACMCtransm} 
Transmission as a function of angular separation for the PIAACMC designs listed in table \ref{tab:PIAACMC}.
}
\end{figure*}

Table \ref{tab:PIAACMC} summarizes the PIAACMC designs discussed in this section. For each design, the circular remapping function was first chosen, and is represented in the table by the parameter $b$, which is the diameter of the focal plane mask used to iteratively compute the generalized prolate function for a circular aperture. A small value of $b$ indicates a weak apodization. The PIAA strength listed in the table is the surface brightness ratio between the brightest and faintest parts of the remapped beam, and is a function of only $b$. This ratio is a good indicator for both the level of distortions of the off-axis PSFs in the intermediate focal plane, and for the difficulty in making the PIAA optics. Current PIAA optics for conventional PIAA coronagraphs have a strength around 100, and any value below 100 therefore corresponds to PIAA optics that can be manufactured to nm-level surface accuracy without technological advances. For PIAA strength values above 100, a hybrid scheme where some of the edge apodization is offloaded to a conventional apodizer should be adopted, at the cost of lower throughput (typically up to 10\% throughput loss) and loss of angular resolution and IWA (by typically up to 5\%).

\begin{figure*}[tb]
\includegraphics[scale=0.66]{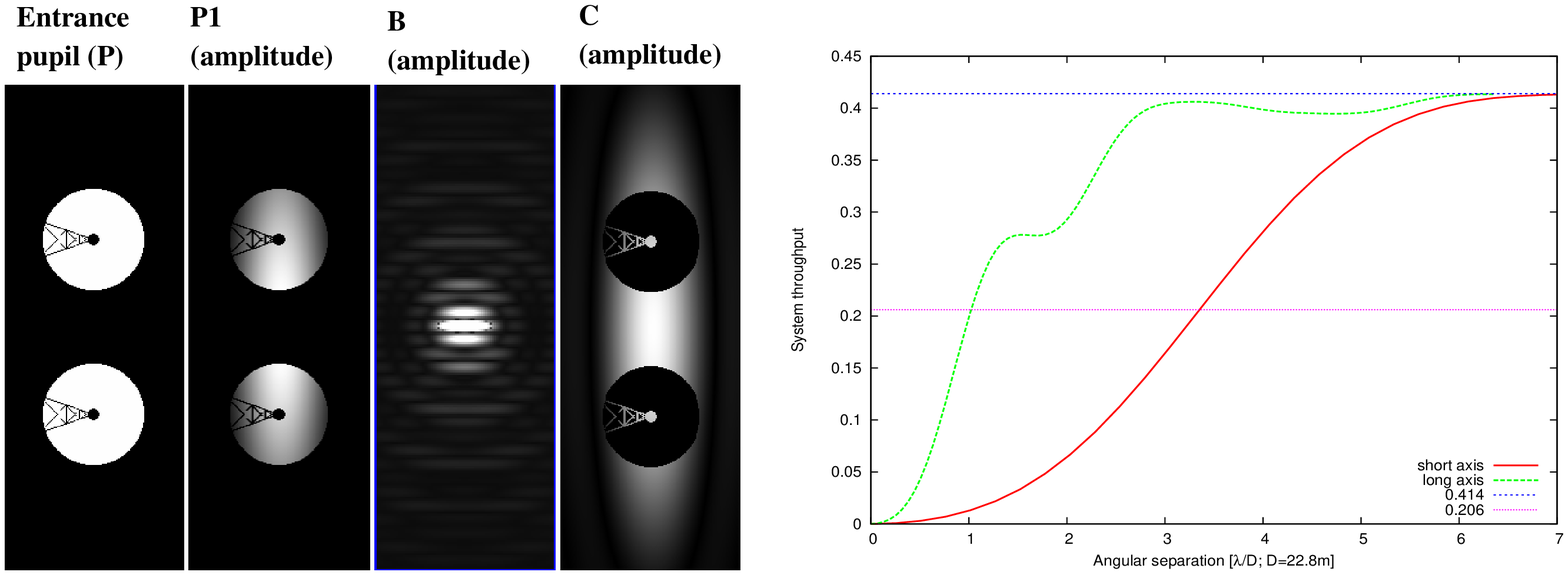} 
\caption{\label{fig:LBT_APCMLC_design} 
Light distribution in selected planes (left) and transmission as a function of angular separation (right) for the LBT APCMLC design. As shown on the right panel, the system throughput at large angular separation, which is equal to the apodizer throughput, is 41.4\%. The half throughput level is also shown, as it defines the coronagraph inner working angle.
}
\end{figure*}

\section{Pupils with strong aspect ratio}
\label{sec:aspectratio}

\subsection{Challenges}
The APCMLC and PIAACMC coronagraphs described in the previous section achieve full starlight suppression by performing, for each point in the output pupil, perfect destructive interference between the light that passes through the circular focal plane mask and the light that passes around it. To offer $\lambda/D$-level inner working angle, these concepts therefore require that the telescope's non coronagraphic point spread function consist of a central diffraction spot within which a disk containing approximately half of the total PSF flux can be drawn, surrounded by other fainter diffractive features (rings, spikes). The examples given in the previous sections (Subaru, GMT, TMT, E-ELT) fulfill this requirement, as these pupil shapes are sufficiently close to a disk.

While very sparse or elongated apertures are not compatible with the APCMLC and PIAACMC concepts as described so far, simple geometric transformations can extend the concepts to a wider range of pupil shapes. The Large Binocular Telescope (LBT) pupil is used in this section as an example of a sparse aperture with a strong aspect ratio: with its two centrally obscured 8.4-m diameter circular subapertures separated by 14.4m (center to center), the LBT pupil has a strong aspect ratio (8.4-m x 22.8-m). The corresponding non-coronagraphic PSF consists of three bright interference fringes within an envelope defined by the single aperture PSF. No circular mask can be drawn within the central bright fringe that contains half of the total PSF flux.

\subsection{Using non-circular focal plane masks}
Stretching the LBT pupil along its narrow direction by a factor four would create a pupil sufficiently close to circular for the APCMLC and PIAACMC concepts as presented above. This stretch is equivalent to using an elliptical focal plane mask, which is four times longer in the direction running along the fringe in the PSF. Figure \ref{fig:LBT_APCMLC_design} shows an APCMLC design for the LBT pupil using an elliptical focal plane mask. The design shown does produce total extinction of an on-axis point source, and its inner working angle is close to 1 $\lambda$/D along the long axis of the pupil (here, D is defined as the diameter of the circle enscribing the pupil, and is equal to 22.8 m), while it is $\approx$ 3.5 $\lambda$/D along the short axis (fundamentally limited by the telescope pupil diffraction along this axis, rather than by the coronagraph). A focal plane mask consisting of three separate zones covering part of the three central fringes may also be adopted to further improve system throughput, although this has not been numerically tested. 

The same elliptical focal plane mask scheme can also applied for the PIAACMC concept on the LBT pupil. Interestingly, the pupil remapping which is part of the PIAACMC concept may be chosen to also bring the two aperture closer to approach the circular pupil case.

The elliptical focal plane mask approach may also be adopted to improve the APCMLC and PIAACMC performance for other non-circular pupil geometries: the focal plane mask shape should ideally be chosen to best match the non-coronagraphic PSF in order to maximize the conventional apodizer's transmission. For example, the generalized prolate function for the Subaru Telescope PIAACMC design \#1 is slightly elongated due to the off-axis spider vanes. This produces a slight mismatch with the circular symmetric remapping function, which is absorbed by the conventional apodizer. Most of the conventional apodizer's light loss (0.1\% total) is due to this mismatch. For this example, using an slightly elliptical focal plane mask would only improve throughput by at most 0.1\% since the pupil is very close to being circular. More importantly, the elliptical focal plane mask may allow high performance operation of the PIAACMC without an apodizer. Adopting a hexagonal shaped focal plane mask would offer similar benefits for hegagonal-shaped pupils such as the one shown in Figures \ref{fig:APCMLCprinciple} and \ref{fig:PIAACMCprinc}.

\subsection{Pupil remapping}

With extremely sparse pupil geometries, the re-design of the focal plane mask geometry may not be sufficient to adapt the pupil shape to the APCMLC and PIAACMC requirements. In this case, geometrical transformation of the sparse entrance pupil into a more compact geometry can be achieved through pupil remapping. This scheme was explored to implement coronagraphy on sparse apertur \citep{2002AA...396..345R, 2002AA...391..379G}, and commonly referred to as the hypertelescope concept.

Even if pupil remapping is not required, it may be useful to improve the APCMLC and PIAACMC system throughput. With sparse apertures, the apodizer becomes less transmissive: for example, the LBT pupil APCMLC design given in this section offers a 41\% throughput, which is significantly less than the $\approx$60\% throughput of comparable APCMLC designs for the Subaru, GMT, TMT and E-ELT pupils. Bringing the LBT subapertures closer together with periscope-like optics would allow for higher throughput in the coronagraph. In order to maintain a good image quality over a wide field of view, the original pupil geometry should be re-created prior to the final imaging focal plane: the compact pupil is only an intermediate step required for efficient removal of the central source's light.

\begin{deluxetable*}{lcccccc}
%\begin{deluxetable}{lcccccc}
% for single column
\tabletypesize{\footnotesize}
\tablecolumns{6} 
\tablewidth{0pc} 
\tablecaption{\label{tab:chrom} Polychromatic performance of a monochromatic PIAACMC design} 
\tablehead{ 
\colhead{Spectral band} & \colhead{Mask thickness} & \colhead{Leakage\tablenotemark{a}} & \multicolumn{3}{c}{Average contrast (0.88 - 3.6 $\lambda/D$)\tablenotemark{b}}\\
 &  &  & \colhead{Best} & \colhead{Worst} & \colhead{Average} 
}
% \multicolumn{3}{c}{
\startdata 
Monochromatic   &  1568.34 nm & 0        &  0       & 0       &    0 \\
2\% band        &  1568.25 nm & 5.29e-5  & 9.56e-11 & 8.31e-7 & 3.10e-7\\
4\% band        &  1568.03 nm & 1.89e-4  & 2.15e-10 & 3.35e-6 & 1.16e-6\\
10\% band       &  1566.54 nm & 1.08e-3  & 2.19e-7  & 2.13e-5 & 7.07e-6\\
20\% band       &  1561.24 nm & 4.30e-3  & 7.68e-7  & 8.57e-5 & 2.80e-5\\
40\% band       &  1543.88 nm & 1.52e-2  & 6.09e-7  & 3.24e-4 & 1.05e-4\\
\enddata 

\tablenotetext{a}{Fraction of the total on-axis source flux that leaks through the coronagraph system, averaged across the spectral band}
\tablenotetext{b}{The average contrast is shown for the best wavelength within the band (left), the worst wavelength within the band (center) and averaged across the band (right)}

\end{deluxetable*} 
%\end{deluxetable} 

\section{Chromaticity}
\label{sec:chrom}

\subsection{Sensitivity to chromatic effects}
All coronagraph systems discussed in this paper were designed for monochromatic light operation. While design of polychromatic APCMLC and PIAACMC systems is outside the scope of this paper (this will be discussed in a future publication), we describe qualitatively in this section how the monochromatic designs perform in broadband light.

Several effects result in a loss of performance in broadband light :
\begin{enumerate}
\item{The physical size of the focal plane mask is adjusted for a single wavelength. While the mask size is independent of wavelength, it should ideally scale linearly with wavelength.}
\item{The phase shift introduced by the mask may vary as a function of wavelength, while it should ideally be constant across the spectral band.}
\item{The transmission of the mask may vary as a function of wavelength, while it should ideally be constant across the band}
\end{enumerate}
The amplitude of the last two effects is a function of how the focal plane mask is manufactured. In this section, we assume that no attempt to achromatize the mask phase shift has been made, and that it consists of a single material deposited on a substrate, with the material thickness adjusted for monochromatic light operation.

The sensitivity to chromatic effects is mostly driven by the focal plane transmission for both APCMLC and PIAACMC systems. Designs with large nearly opaque focal plane masks are more tolerant to chromatic effects, since the mask's role becomes close to a simple light block, and he mask size relative to the on-axis source image increases. To illustrate broadband performance, we adopt in the next section a monochromatic PIAACMC design with partial ($0<|t|<1$) focal plane mask transmission. 

\subsection{Example: PIAACMC design for a centrally obscured pupil}
We consider the PIAACMC design 2 for the Subaru Telescope pupil in this section. It is assumed that the focal plane mask size is optimized for monochromatic light at $\lambda = 1.65 \mu m$, and that the mask is a disk of material ($SiO_2$). The mask transmission is fixed to the ideal monochromatic value, and is not assumed to change with wavelength. Several scenarios are considered: monochromatic, 2\%, 4\%, 10\%, 20\% and 40\% wide bands (all centered at 1.65 $\mu$m).

The mask thickness is a free parameter, and is adjusted for each case to yield the best broadband on-axis extinction, as measured by the total light in the final focal plane. Results are shown in table \ref{tab:chrom}. The last 3 columns of the table show spatially averaged contrast values between the coronagraph's inner working angle ($0.88 \lambda / D$) and $3.6 \lambda/D$.

\begin{figure*}[tb]
\includegraphics[scale=0.85]{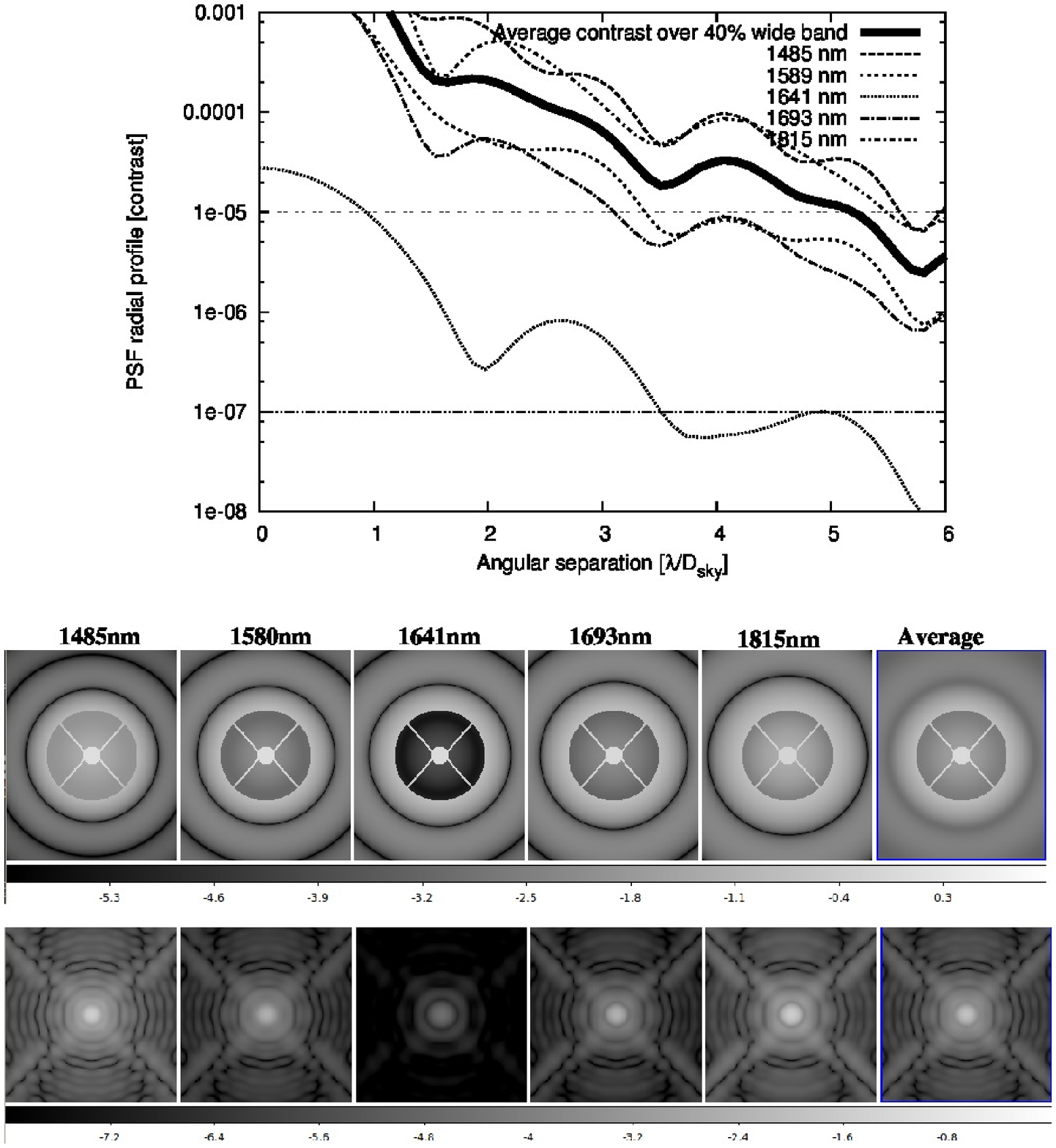} 
\caption{\label{fig:chrom} 
Polychromatic performance across a 40\% wide band for a monochromatic PIAACMC designed for $\lambda=$1650nm for the Subaru Tescope pupil. PSF radial profiles are shown at different wavelengths (top). Pupil plane amplitude (center) and focal plane intensity (bottom) images are shown in log contrast scale.
}
\end{figure*}

This particular design delivers better than 1e-4 averaged raw contrast in a 20\% wide specral band, and is therefore valuable for ground-based use behind adaptive optics. Pupil and focal plane images and contrast radial profiles are shown in Figure \ref{fig:chrom} across a 40\% wide band centered at $1.65 \mu m$, illustrating that raw contrast is best at the center of the band.

\section{Conclusions}
\label{sec:conclusion}

The APCMLC and PIAACMC concepts, previously proposed for unobstructed circular apertures, are also applicable to telescopes with arbitrary pupil shapes. Their performance is largely unaffected by aperture shape, and full throughput low-IWA coronagraphy is therefore theoretically possible on any pupil shape with the PIAACMC. The demonstration that the coronagraph with the highest known theoretical performance can be applied on any pupil may remove the requirement that a future space-based exoplanet direct imaging mission should use an off-axis telescope. On ground-based telescopes, which adopt optical designs which are generally not driven by coronagraphy, high efficiency coronagraphy at and within 1 $\lambda/D$ is possible, potentially allowing direct imaging of habitable planets around nearby M and K type main sequence stars for which the planet-to-star contrast is favorable but the angular separation is extremely challenging and requires $\approx \lambda/D$ IWA even on a 30-m class telescope.

Manufacturing and implementation challenges have not been addressed in this paper. Manufacturing an achromatic focal plane mask for the APCMLC or PIAACMC is challenging, as its size should scale linearly with wavelength, and its complex amplitude transmission should be achromatic. Similar challenges have been previously addressed for other coronagraphs \citep{2011ApJ...729..144S}, using carefully designed multilayer coatings of variable thickness and/or sub-$\lambda/D$ sized features optimized to produced the required chromatic dependence within the geometric pupil \citep{2003AA...403..369S,2012AA...538A..55N}. The PIAA optics required for the PIAACMC are however not as challenging to manufacture as PIAA optics previously made for hard edged opaque focal plane masks, because the PIAACMC's entrance apodization is milder. As any high performance low IWA coronagraph, the PIAACMC performance is highly sensitive to residual wavefront errors, which must be actively sensed and controlled. The PIAACMC's high throughput is an asset for achieving the required wavefront quality, as wavefront sensing can be performed rapidly, using all incoming light. 

Small IWA high contrast coronagraphy requires exquisite control of tip-tilt and low order wavefront errors. The central star angular size may also impose limits on the achievable performance \citep{2009ApJ...702..672C}. These issues have not been addressed or quantified in this paper, but may drive the optimal coronagraph design for a particular scientific application. We note that both sub-$\lambda/D$ IWA coronagraphs designs described in this paper can also be designed for IWA equal to or larger than $\lambda$/D if required, offering lower performance but improved resilience against pointing errors and stellar angular size. While the APCMLC design with larger IWA has a lower throughput (due to the stronger apodization), for the PIAACMC, the large-IWA designs maintain full throughput and total on-axis extinction, offering a wide range of practical high performance coronagraphic options. $\:$ $\:$ $\:$ $\:$ $\:$ $\:$              -           

\newpage
%\acknowledgments
%This work is funded by the 

\bibliography{ms}

\begin{thebibliography}{45}
\expandafter\ifx\csname natexlab\endcsname\relax\def\natexlab#1{#1}\fi

\bibitem[{{Aime} {et~al.}(2002){Aime}, {Soummer}, \&
  {Ferrari}}]{2002A&A...389..334A}
{Aime}, C., {Soummer}, R., \& {Ferrari}, A. 2002, \aap, 389, 334

\bibitem[{{Baudoz} {et~al.}(2000){Baudoz}, {Rabbia}, \&
  {Gay}}]{2000A&AS..141..319B}
{Baudoz}, P., {Rabbia}, Y., \& {Gay}, J. 2000, \aaps, 141, 319

\bibitem[{{Beuzit} {et~al.}(2008){Beuzit}, {Feldt}, {Dohlen}, {Mouillet},
  {Puget}, {Wildi}, {Abe}, {Antichi}, {Baruffolo}, {Baudoz}, {Boccaletti},
  {Carbillet}, {Charton}, {Claudi}, {Downing}, {Fabron}, {Feautrier},
  {Fedrigo}, {Fusco}, {Gach}, {Gratton}, {Henning}, {Hubin}, {Joos}, {Kasper},
  {Langlois}, {Lenzen}, {Moutou}, {Pavlov}, {Petit}, {Pragt}, {Rabou}, {Rigal},
  {Roelfsema}, {Rousset}, {Saisse}, {Schmid}, {Stadler}, {Thalmann}, {Turatto},
  {Udry}, {Vakili}, \& {Waters}}]{2008SPIE.7014E..41B}
{Beuzit}, J.-L., {Feldt}, M., {Dohlen}, K., {Mouillet}, D., {Puget}, P.,
  {Wildi}, F., {Abe}, L., {Antichi}, J., {Baruffolo}, A., {Baudoz}, P.,
  {Boccaletti}, A., {Carbillet}, M., {Charton}, J., {Claudi}, R., {Downing},
  M., {Fabron}, C., {Feautrier}, P., {Fedrigo}, E., {Fusco}, T., {Gach}, J.-L.,
  {Gratton}, R., {Henning}, T., {Hubin}, N., {Joos}, F., {Kasper}, M.,
  {Langlois}, M., {Lenzen}, R., {Moutou}, C., {Pavlov}, A., {Petit}, C.,
  {Pragt}, J., {Rabou}, P., {Rigal}, F., {Roelfsema}, R., {Rousset}, G.,
  {Saisse}, M., {Schmid}, H.-M., {Stadler}, E., {Thalmann}, C., {Turatto}, M.,
  {Udry}, S., {Vakili}, F., \& {Waters}, R. 2008, in Society of Photo-Optical
  Instrumentation Engineers (SPIE) Conference Series, Vol. 7014, Society of
  Photo-Optical Instrumentation Engineers (SPIE) Conference Series

\bibitem[{{Carlotti} {et~al.}(2011){Carlotti}, {Vanderbei}, \&
  {Kasdin}}]{2011arXiv1108.4050C}
{Carlotti}, A., {Vanderbei}, R., \& {Kasdin}, N.~J. 2011, ArXiv e-prints

\bibitem[{{Carson} {et~al.}(2013){Carson}, {Thalmann}, {Janson}, {Kozakis},
  {Bonnefoy}, {Biller}, {Schlieder}, {Currie}, {McElwain}, {Goto}, {Henning},
  {Brandner}, {Feldt}, {Kandori}, {Kuzuhara}, {Stevens}, {Wong}, {Gainey},
  {Fukagawa}, {Kuwada}, {Brandt}, {Kwon}, {Abe}, {Egner}, {Grady}, {Guyon},
  {Hashimoto}, {Hayano}, {Hayashi}, {Hayashi}, {Hodapp}, {Ishii}, {Iye},
  {Knapp}, {Kudo}, {Kusakabe}, {Matsuo}, {Miyama}, {Morino}, {Moro-Martin},
  {Nishimura}, {Pyo}, {Serabyn}, {Suto}, {Suzuki}, {Takami}, {Takato},
  {Terada}, {Tomono}, {Turner}, {Watanabe}, {Wisniewski}, {Yamada}, {Takami},
  {Usuda}, \& {Tamura}}]{2013ApJ...763L..32C}
{Carson}, J., {Thalmann}, C., {Janson}, M., {Kozakis}, T., {Bonnefoy}, M.,
  {Biller}, B., {Schlieder}, J., {Currie}, T., {McElwain}, M., {Goto}, M.,
  {Henning}, T., {Brandner}, W., {Feldt}, M., {Kandori}, R., {Kuzuhara}, M.,
  {Stevens}, L., {Wong}, P., {Gainey}, K., {Fukagawa}, M., {Kuwada}, Y.,
  {Brandt}, T., {Kwon}, J., {Abe}, L., {Egner}, S., {Grady}, C., {Guyon}, O.,
  {Hashimoto}, J., {Hayano}, Y., {Hayashi}, M., {Hayashi}, S., {Hodapp}, K.,
  {Ishii}, M., {Iye}, M., {Knapp}, G., {Kudo}, T., {Kusakabe}, N., {Matsuo},
  T., {Miyama}, S., {Morino}, J., {Moro-Martin}, A., {Nishimura}, T., {Pyo},
  T., {Serabyn}, E., {Suto}, H., {Suzuki}, R., {Takami}, M., {Takato}, N.,
  {Terada}, H., {Tomono}, D., {Turner}, E., {Watanabe}, M., {Wisniewski}, J.,
  {Yamada}, T., {Takami}, H., {Usuda}, T., \& {Tamura}, M. 2013, \apjl, 763,
  L32

\bibitem[{{Crepp} {et~al.}(2009){Crepp}, {Mahadevan}, \&
  {Ge}}]{2009ApJ...702..672C}
{Crepp}, J.~R., {Mahadevan}, S., \& {Ge}, J. 2009, \apj, 702, 672

\bibitem[{{Crepp} {et~al.}(2011){Crepp}, {Pueyo}, {Brenner}, {Oppenheimer},
  {Zimmerman}, {Hinkley}, {Parry}, {King}, {Vasisht}, {Beichman},
  {Hillenbrand}, {Dekany}, {Shao}, {Burruss}, {Roberts}, {Bouchez}, {Roberts},
  \& {Soummer}}]{2011ApJ...729..132C}
{Crepp}, J.~R., {Pueyo}, L., {Brenner}, D., {Oppenheimer}, B.~R., {Zimmerman},
  N., {Hinkley}, S., {Parry}, I., {King}, D., {Vasisht}, G., {Beichman}, C.,
  {Hillenbrand}, L., {Dekany}, R., {Shao}, M., {Burruss}, R., {Roberts}, L.~C.,
  {Bouchez}, A., {Roberts}, J., \& {Soummer}, R. 2011, \apj, 729, 132

\bibitem[{{Guyon}(2002)}]{guyonPhD}
{Guyon}, O. 2002, PhD thesis

\bibitem[{{Guyon}(2003)}]{2003AA...404..379G}
---. 2003, \aap, 404, 379

\bibitem[{{Guyon} {et~al.}(2010){Guyon}, {Martinache}, {Belikov}, \&
  {Soummer}}]{2010ApJS..190..220G}
{Guyon}, O., {Martinache}, F., {Belikov}, R., \& {Soummer}, R. 2010, \apjs,
  190, 220

\bibitem[{{Guyon} {et~al.}(1999){Guyon}, {Roddier}, {Graves}, {Roddier},
  {Cuevas}, {Espejo}, {Gonzalez}, {Martinez}, {Bisiacchi}, \&
  {Vuntesmeri}}]{1999PASP..111.1321G}
{Guyon}, O., {Roddier}, C., {Graves}, J.~E., {Roddier}, F., {Cuevas}, S.,
  {Espejo}, C., {Gonzalez}, S., {Martinez}, A., {Bisiacchi}, G., \&
  {Vuntesmeri}, V. 1999, \pasp, 111, 1321

\bibitem[{{Guyon} \& {Roddier}(2002)}]{2002AA...391..379G}
{Guyon}, O. \& {Roddier}, F. 2002, \aap, 391, 379

\bibitem[{{Guyon} \& {Roddier}(2000)}]{2000SPIE.4006..377G}
{Guyon}, O. \& {Roddier}, F.~J. 2000, in Society of Photo-Optical
  Instrumentation Engineers (SPIE) Conference Series, Vol. 4006, Society of
  Photo-Optical Instrumentation Engineers (SPIE) Conference Series, ed.
  {P.~L{\'e}na \& A.~Quirrenbach}, 377--387

\bibitem[{{Guyon} \& {Shao}(2006)}]{2006PASP..118..860G}
{Guyon}, O. \& {Shao}, M. 2006, \pasp, 118, 860

\bibitem[{{Hinkley} {et~al.}(2011){Hinkley}, {Oppenheimer}, {Zimmerman},
  {Brenner}, {Parry}, {Crepp}, {Vasisht}, {Ligon}, {King}, {Soummer},
  {Sivaramakrishnan}, {Beichman}, {Shao}, {Roberts}, {Bouchez}, {Dekany},
  {Pueyo}, {Roberts}, {Lockhart}, {Zhai}, {Shelton}, \&
  {Burruss}}]{2011PASP..123...74H}
{Hinkley}, S., {Oppenheimer}, B.~R., {Zimmerman}, N., {Brenner}, D., {Parry},
  I.~R., {Crepp}, J.~R., {Vasisht}, G., {Ligon}, E., {King}, D., {Soummer}, R.,
  {Sivaramakrishnan}, A., {Beichman}, C., {Shao}, M., {Roberts}, L.~C.,
  {Bouchez}, A., {Dekany}, R., {Pueyo}, L., {Roberts}, J.~E., {Lockhart}, T.,
  {Zhai}, C., {Shelton}, C., \& {Burruss}, R. 2011, \pasp, 123, 74

\bibitem[{{Krist} {et~al.}(2009){Krist}, {Balasubramanian}, {Beichman},
  {Echternach}, {Green}, {Liewer}, {Muller}, {Serabyn}, {Shaklan}, {Trauger},
  {Wilson}, {Horner}, {Mao}, {Somerstein}, {Vasudevan}, {Kelly}, \&
  {Rieke}}]{2009SPIE.7440E..28K}
{Krist}, J.~E., {Balasubramanian}, K., {Beichman}, C.~A., {Echternach}, P.~M.,
  {Green}, J.~J., {Liewer}, K.~M., {Muller}, R.~E., {Serabyn}, E., {Shaklan},
  S.~B., {Trauger}, J.~T., {Wilson}, D.~W., {Horner}, S.~D., {Mao}, Y.,
  {Somerstein}, S.~F., {Vasudevan}, G., {Kelly}, D.~M., \& {Rieke}, M.~J. 2009,
  in Society of Photo-Optical Instrumentation Engineers (SPIE) Conference
  Series, Vol. 7440, Society of Photo-Optical Instrumentation Engineers (SPIE)
  Conference Series

\bibitem[{{Kuchner} \& {Traub}(2002)}]{2002ApJ...570..900K}
{Kuchner}, M.~J. \& {Traub}, W.~A. 2002, \apj, 570, 900

\bibitem[{{Lagrange} {et~al.}(2010){Lagrange}, {Bonnefoy}, {Chauvin}, {Apai},
  {Ehrenreich}, {Boccaletti}, {Gratadour}, {Rouan}, {Mouillet}, {Lacour}, \&
  {Kasper}}]{2010Sci...329...57L}
{Lagrange}, A.-M., {Bonnefoy}, M., {Chauvin}, G., {Apai}, D., {Ehrenreich}, D.,
  {Boccaletti}, A., {Gratadour}, D., {Rouan}, D., {Mouillet}, D., {Lacour}, S.,
  \& {Kasper}, M. 2010, Science, 329, 57

\bibitem[{{Levine} {et~al.}(2009){Levine}, {Lisman}, {Shaklan}, {Kasting},
  {Traub}, {Alexander}, {Angel}, {Blaurock}, {Brown}, {Brown}, {Burrows},
  {Clampin}, {Cohen}, {Content}, {Dewell}, {Dumont}, {Egerman}, {Ferguson},
  {Ford}, {Greene}, {Guyon}, {Hammel}, {Heap}, {Ho}, {Horner}, {Hunyadi},
  {Irish}, {Jackson}, {Kasdin}, {Kissil}, {Krim}, {Kuchner}, {Kwack}, {Lillie},
  {Lin}, {Liu}, {Marchen}, {Marley}, {Meadows}, {Mosier}, {Mouroulis},
  {Noecker}, {Ohl}, {Oppenheimer}, {Pitman}, {Ridgway}, {Sabatke}, {Seager},
  {Shao}, {Smith}, {Soummer}, {Stapelfeldt}, {Tenerell}, {Trauger}, \&
  {Vanderbei}}]{2009arXiv0911.3200L}
{Levine}, M., {Lisman}, D., {Shaklan}, S., {Kasting}, J., {Traub}, W.,
  {Alexander}, J., {Angel}, R., {Blaurock}, C., {Brown}, M., {Brown}, R.,
  {Burrows}, C., {Clampin}, M., {Cohen}, E., {Content}, D., {Dewell}, L.,
  {Dumont}, P., {Egerman}, R., {Ferguson}, H., {Ford}, V., {Greene}, J.,
  {Guyon}, O., {Hammel}, H., {Heap}, S., {Ho}, T., {Horner}, S., {Hunyadi}, S.,
  {Irish}, S., {Jackson}, C., {Kasdin}, J., {Kissil}, A., {Krim}, M.,
  {Kuchner}, M., {Kwack}, E., {Lillie}, C., {Lin}, D., {Liu}, A., {Marchen},
  L., {Marley}, M., {Meadows}, V., {Mosier}, G., {Mouroulis}, P., {Noecker},
  M., {Ohl}, R., {Oppenheimer}, B., {Pitman}, J., {Ridgway}, S., {Sabatke}, E.,
  {Seager}, S., {Shao}, M., {Smith}, A., {Soummer}, R., {Stapelfeldt}, K.,
  {Tenerell}, D., {Trauger}, J., \& {Vanderbei}, R. 2009, ArXiv e-prints

\bibitem[{{Lloyd} {et~al.}(2003){Lloyd}, {Gavel}, {Graham}, {Hodge},
  {Sivaramakrishnan}, \& {Voit}}]{2003SPIE.4860..171L}
{Lloyd}, J.~P., {Gavel}, D.~T., {Graham}, J.~R., {Hodge}, P.~E.,
  {Sivaramakrishnan}, A., \& {Voit}, G.~M. 2003, in Society of Photo-Optical
  Instrumentation Engineers (SPIE) Conference Series, Vol. 4860, Society of
  Photo-Optical Instrumentation Engineers (SPIE) Conference Series, ed.
  {A.~B.~Schultz}, 171--181

\bibitem[{{Lozi} {et~al.}(2009){Lozi}, {Martinache}, \&
  {Guyon}}]{2009PASP..121.1232L}
{Lozi}, J., {Martinache}, F., \& {Guyon}, O. 2009, \pasp, 121, 1232

\bibitem[{{Macintosh} {et~al.}(2008){Macintosh}, {Graham}, {Palmer}, {Doyon},
  {Dunn}, {Gavel}, {Larkin}, {Oppenheimer}, {Saddlemyer}, {Sivaramakrishnan},
  {Wallace}, {Bauman}, {Erickson}, {Marois}, {Poyneer}, \&
  {Soummer}}]{2008SPIE.7015E..31M}
{Macintosh}, B.~A., {Graham}, J.~R., {Palmer}, D.~W., {Doyon}, R., {Dunn}, J.,
  {Gavel}, D.~T., {Larkin}, J., {Oppenheimer}, B., {Saddlemyer}, L.,
  {Sivaramakrishnan}, A., {Wallace}, J.~K., {Bauman}, B., {Erickson}, D.~A.,
  {Marois}, C., {Poyneer}, L.~A., \& {Soummer}, R. 2008, in Society of
  Photo-Optical Instrumentation Engineers (SPIE) Conference Series, Vol. 7015,
  Society of Photo-Optical Instrumentation Engineers (SPIE) Conference Series

\bibitem[{{Marois} {et~al.}(2008){Marois}, {Macintosh}, {Barman}, {Zuckerman},
  {Song}, {Patience}, {Lafreni{\`e}re}, \& {Doyon}}]{2008Sci...322.1348M}
{Marois}, C., {Macintosh}, B., {Barman}, T., {Zuckerman}, B., {Song}, I.,
  {Patience}, J., {Lafreni{\`e}re}, D., \& {Doyon}, R. 2008, Science, 322, 1348

\bibitem[{{Martinache} \& {Guyon}(2009)}]{2009SPIE.7440E..20M}
{Martinache}, F. \& {Guyon}, O. 2009, in Society of Photo-Optical
  Instrumentation Engineers (SPIE) Conference Series, Vol. 7440, Society of
  Photo-Optical Instrumentation Engineers (SPIE) Conference Series

\bibitem[{{Martinez}(2010)}]{2010AA...519A..61M}
{Martinez}, P. 2010, \aap, 519, A61

\bibitem[{{Martinez} {et~al.}(2007){Martinez}, {Boccaletti}, {Kasper},
  {Baudoz}, \& {Cavarroc}}]{2007AA...474..671M}
{Martinez}, P., {Boccaletti}, A., {Kasper}, M., {Baudoz}, P., \& {Cavarroc}, C.
  2007, \aap, 474, 671

\bibitem[{{Martinez} {et~al.}(2008){Martinez}, {Boccaletti}, {Kasper},
  {Cavarroc}, {Yaitskova}, {Fusco}, \& {V{\'e}rinaud}}]{2008AA...492..289M}
{Martinez}, P., {Boccaletti}, A., {Kasper}, M., {Cavarroc}, C., {Yaitskova},
  N., {Fusco}, T., \& {V{\'e}rinaud}, C. 2008, \aap, 492, 289

\bibitem[{{Martinez} {et~al.}(2010){Martinez}, {Dorrer}, {Kasper},
  {Boccaletti}, \& {Dohlen}}]{2010AA...520A.110M}
{Martinez}, P., {Dorrer}, C., {Kasper}, M., {Boccaletti}, A., \& {Dohlen}, K.
  2010, \aap, 520, A110

\bibitem[{{Mawet} {et~al.}(2011){Mawet}, {Serabyn}, {Wallace}, \&
  {Pueyo}}]{2011OptL...36.1506M}
{Mawet}, D., {Serabyn}, E., {Wallace}, J.~K., \& {Pueyo}, L. 2011, Optics
  Letters, 36, 1506

\bibitem[{{Murakami} \& {Baba}(2005)}]{2005PASP..117..295M}
{Murakami}, N. \& {Baba}, N. 2005, \pasp, 117, 295

\bibitem[{{N'diaye} {et~al.}(2010){N'diaye}, {Dohlen}, {Cuevas}, {Lanzoni},
  {Chemla}, {Chaumont}, {Soummer}, \& {Griffiths}}]{2010AA...509A...8N}
{N'diaye}, M., {Dohlen}, K., {Cuevas}, S., {Lanzoni}, P., {Chemla}, F.,
  {Chaumont}, C., {Soummer}, R., \& {Griffiths}, E.~T. 2010, \aap, 509, A8

\bibitem[{{N'diaye} {et~al.}(2012){N'diaye}, {Dohlen}, {Cuevas}, {Soummer},
  {S{\'a}nchez-P{\'e}rez}, \& {Zamkotsian}}]{2012AA...538A..55N}
{N'diaye}, M., {Dohlen}, K., {Cuevas}, S., {Soummer}, R.,
  {S{\'a}nchez-P{\'e}rez}, C., \& {Zamkotsian}, F. 2012, \aap, 538, A55

\bibitem[{{Riaud} {et~al.}(2002){Riaud}, {Boccaletti}, {Gillet}, {Schneider},
  {Labeyrie}, {Arnold}, {Baudrand}, {Lardi{\`e}}, {Dejonghe}, \&
  {Borkowski}}]{2002AA...396..345R}
{Riaud}, P., {Boccaletti}, A., {Gillet}, S., {Schneider}, J., {Labeyrie}, A.,
  {Arnold}, L., {Baudrand}, J., {Lardi{\`e}}, O., {Dejonghe}, J., \&
  {Borkowski}, V. 2002, \aap, 396, 345

\bibitem[{{Roddier} \& {Roddier}(1997)}]{1997PASP..109..815R}
{Roddier}, F. \& {Roddier}, C. 1997, \pasp, 109, 815

\bibitem[{{Serabyn} {et~al.}(2007){Serabyn}, {Wallace}, {Troy}, {Mennesson},
  {Haguenauer}, {Gappinger}, \& {Burruss}}]{2007ApJ...658.1386S}
{Serabyn}, E., {Wallace}, K., {Troy}, M., {Mennesson}, B., {Haguenauer}, P.,
  {Gappinger}, R., \& {Burruss}, R. 2007, \apj, 658, 1386

\bibitem[{{Sivaramakrishnan} \& {Lloyd}(2005)}]{2005ApJ...633..528S}
{Sivaramakrishnan}, A. \& {Lloyd}, J.~P. 2005, \apj, 633, 528

\bibitem[{{Sivaramakrishnan} \& {Yaitskova}(2005)}]{2005ApJ...626L..65S}
{Sivaramakrishnan}, A. \& {Yaitskova}, N. 2005, \apjl, 626, L65

\bibitem[{{Soummer}(2005)}]{2005ApJ...618L.161S}
{Soummer}, R. 2005, \apjl, 618, L161

\bibitem[{{Soummer} {et~al.}(2003{\natexlab{a}}){Soummer}, {Aime}, \&
  {Falloon}}]{2003AA...397.1161S}
{Soummer}, R., {Aime}, C., \& {Falloon}, P.~E. 2003{\natexlab{a}}, \aap, 397,
  1161

\bibitem[{{Soummer} {et~al.}(2003{\natexlab{b}}){Soummer}, {Dohlen}, \&
  {Aime}}]{2003AA...403..369S}
{Soummer}, R., {Dohlen}, K., \& {Aime}, C. 2003{\natexlab{b}}, \aap, 403, 369

\bibitem[{{Soummer} {et~al.}(2009){Soummer}, {Pueyo}, {Ferrari}, {Aime},
  {Sivaramakrishnan}, \& {Yaitskova}}]{2009ApJ...695..695S}
{Soummer}, R., {Pueyo}, L., {Ferrari}, A., {Aime}, C., {Sivaramakrishnan}, A.,
  \& {Yaitskova}, N. 2009, \apj, 695, 695

\bibitem[{{Soummer} {et~al.}(2011){Soummer}, {Sivaramakrishnan}, {Pueyo},
  {Macintosh}, \& {Oppenheimer}}]{2011ApJ...729..144S}
{Soummer}, R., {Sivaramakrishnan}, A., {Pueyo}, L., {Macintosh}, B., \&
  {Oppenheimer}, B.~R. 2011, \apj, 729, 144

\bibitem[{{Tanaka} {et~al.}(2006){Tanaka}, {Enya}, {Abe}, {Nakagawa}, \&
  {Kataza}}]{2006PASJ...58..627T}
{Tanaka}, S., {Enya}, K., {Abe}, L., {Nakagawa}, T., \& {Kataza}, H. 2006,
  \pasj, 58, 627

\bibitem[{{Thomas} {et~al.}(2011){Thomas}, {Soummer}, {Dillon}, {Macintosh},
  {Gavel}, \& {Sivaramakrishnan}}]{2011AJ....142..119T}
{Thomas}, S.~J., {Soummer}, R., {Dillon}, D., {Macintosh}, B., {Gavel}, D., \&
  {Sivaramakrishnan}, A. 2011, \aj, 142, 119

\bibitem[{{Trauger} \& {Traub}(2007)}]{2007Natur.446..771T}
{Trauger}, J.~T. \& {Traub}, W.~A. 2007, \nat, 446, 771

\end{thebibliography}

\end{document}